\documentclass{aa}  

\usepackage{graphicx}
\usepackage{txfonts}
\usepackage{bm}

\begin{document}

   \title{Stellar dating using chemical clocks and Bayesian inference}

   \author{A. Moya
          \inst{1,2,3,4}
          \and
          L.~M. Sarro\inst{5}
          \and
          E. Delgado-Mena\inst{6}
          \and
2          W.~J. Chaplin\inst{3,4}
          \and V. Adibekyan\inst{6,7}
          \and S. Blanco-Cuaresma\inst{8}
          }

   \institute{
   Departament d’Astronomia i Astrofísica, Universitat de València, C. Dr. Moliner 50, 46100 Burjassot, Spain
   \and
   Electrical Engineering, Electronics, Automation and Applied Physics Department, E.T.S.I.D.I, Polytechnic University of Madrid (UPM), Madrid 28012, Spain 
   \and
    School of Physics and Astronomy, University of Birmingham, Edgbaston, Birmingham, B15 2TT, UK
    \and
    Stellar Astrophysics Centre, Department of Physics and Astronomy, Aarhus University, Ny Munkegade 120, DK-8000 Aarhus C, Denmark
    \and
    Departamento de Inteligencia Artificial, ETSI Informática, UNED, Juan del Rosal, E-16 28040 Madrid, Spain
    \and
    Instituto de Astrof\'isica e Ci\^encias do Espa\c{c}o, Universidade do Porto, CAUP, Rua das Estrelas, 4150-762 Porto, Portugal
    \and
    Departamento de F\'{\i}sica e Astronomia, Faculdade de Ci\^encias, Universidade do Porto, Rua do Campo  Alegre, 4169-007 Porto, Portugal
    \and
    Harvard-Smithsonian Center for Astrophysics, 60 Garden Street, Cambridge, MA 02138, USA
             }

   \date{}

 
  \abstract
   {Dating stars is a major challenge with a deep impact on many astrophysical fields. One of the most promising techniques for this is using chemical abundances. Recent space- and ground-based facilities have improved the quantity of stars with accurate observations. This has opened the door for using Bayesian inference tools to maximise the information we can extract from them.}
   {Our aim is to present accurate and reliable stellar age estimates of FGK stars using chemical abundances and stellar parameters.}
   {We used one of the most flexible Bayesian inference techniques (hierarchical Bayesian models) to exceed current possibilities in the use of chemical abundances for stellar dating. Our model is a data-driven model. We used a training set that has been presented in the literature with ages estimated with isochrones and accurate stellar abundances and general characteristics. The core of the model is a prescription of certain abundance ratios as linear combinations of stellar properties including age. We gathered four different testing sets to assess the accuracy, precision, and limits of our model. We also trained a model using chemical abundances alone.}
   {We found that our age estimates and those coming from asteroseismology, other accurate sources, and also with ten Gaia benchmark stars agree well. The mean absolute difference of our estimates compared with those used as reference is 0.9 Ga, with a mean difference of 0.01 Ga. When using open clusters, we reached a very good agreement for Hyades, NGC 2632, Ruprecht 147, and IC4651. We also found outliers that are a reflection of chemical peculiarities and/or stars at the limit of the validity ranges of the training set. The model that only uses chemical abundances shows slightly worse mean absolute difference (1.18 Ga) and mean difference (-0.12 Ga).}
   {}


   \maketitle
%

\section{Introduction}

Among all the stellar characteristics, age is one of the most difficult variables to measure because it cannot be directly observed. It must be inferred using diverse methods. \citet{Soderblom10,Soderblom15} presented a summary of most of these techniques and proposed their classification into five groups: fundamental, empirical, semi-empirical, statistical, and modelling.

One of these techniques is the so-called chemical clocks (CCs) method. It exploits the fact of the chemical evolution of the Galaxy. This chemical evolution is a consequence of the dependence of the stellar fusion and thermonuclear reactions operating in stellar interiors (and the atomic elements thereby created) and the stellar evolution depending on the stellar mass, that is, the more massive the star, the faster its evolution (of the order of Ma\footnote{\url{https://www.iau.org/publications/proceedings_rules/units/}} for the most massive ones, with masses higher than 8M$_\odot$). Low-mass stars (M<8M$_\odot$) evolve far more slowly, of the order of even Ga\footnotemark[1] , and the chemical elements that are created are different in general \citep{elements_origin}.

We can assume that the current stellar surface abundances of certain elements are those of the original cloud from which the star was born, that is, stars act as fossil relics in terms of chemical composition. This can be verified using stellar structure and evolution models, where only slight surface abundance variations are predicted during stellar evolution \citep{dotter, gavel}. Therefore, we can use these abundances to estimate the age of the star. This concept has been used in recent years to propose a few chemical abundance ratios for which this evolution is especially clear. The different contribution to the chemical evolution of the Galaxy of supernovae of types II and Ia (SNe II and SNe Ia, respectively) and low-mass asymptotic giant branch stars (AGB) opens the door to the stellar dating using certain surface chemical abundances \citep{Nissen16}. The work by \citet{Silva12} was the first to explore the relation with age of abundance ratios of Y or Sr over Mg, Al, or Zn. More recently, \citet{Nissen15,Nissen16} found that ratios of [Y/Mg], [Y/Al], or [Al/Mg] are precise age indicators in the case of solar twin stars. These are the so-called CCs and have been studied in other samples of solar twins \citep{Spina16,Tucci16}, in a larger sample of stars within the AMBRE project \citep{Titarenko19, Santos21}, and recently in a number of papers \citep[e.g.][]{Casamiquela21, Rebassa21, Taut21, Espinoza21, Morel20, Casali}. Moreover, the application of these CCs to solar twin stars was cross-checked using stars dated by asteroseismology \citep{Nissen17,Jofre20}. However, \citet{Feltzing17} and \citet{Elisa19} (DM19) reported that when stars of different metallicities and/or effective temperatures are included, these simple correlations are no longer valid. DM19 defined up to ten CCs presenting different linear expressions for stellar age estimations involving different observables and different numbers of dimensions, and extended the validity range of these ratios beyond solar twins, increasing the utility of these ratios. This dependence of the age - CC relation on stellar metallicity has been confirmed for some CCs and stars in the Galactic disk by \cite{Casali}. However, these authors and also \cite{Magrini21} and \cite{Katz21} cautioned that that the CCs might not be applicable for all the stars in the Galaxy, in particular not for those in the inner disk. That is, there is a dependence of the age versus CC relations on the Galactocentric distance. In particular, \cite{Casali} and \cite{Magrini21} found that the 2D relation between stellar age, a CC, and [Fe/H] does not map all the Galaxy.

On the other hand, CCs such as Li \citep{Llorente} or the ratio [C/N] \citep{Casali19, Jofre21} are known in the literature. These CCs are based on stellar evolution and they are not considered for this work.

\citet{Morel20} recently extended the use of asteroseismic ages beyond solar twins using the Kepler Legacy data-base to explore these CCs, in particular those presented in DM19. They also confirmed these relations and reported that seismic ages and ages from 3D formulas agree well and that the differences were below typical error levels. Nevertheless, they also reported that CC ages are systematically younger than seismic ages.

Recent studies have explored the use of machine-learning techniques in this context. \citet{Sharma20}, using the GALAH survey, proposed a number of 2D relations where the stellar age is estimated as a function of the stellar metallicity (Fe/H) and the abundance of different elements over iron. Stellar ages for the training set were obtained using evolutionary models. \citet{Hayden20} used this work to improve age estimates using one of the most efficient tools for stacking. They combined a number of weak estimators such as those coming from each 2D relation to construct a strong estimator using the XGBoost algorithm.

Chemical clocks are useful not only for stellar dating. They can also be used to distinguish Galactic events such as the existence of two episodes of accretion of gas onto the Galactic disk with an episode of star formation in between \citep{Nissen20}, and to understand the timescale of different nuclear processes in the Galaxy \citep{Jofre20} or in nearby galaxies \citep{Skuladottir}.

In this paper, we take advantage of the extraordinary data set presented in DM19 and the power of machine-learning techniques to go a step further. We present the best possible age estimates using CCs and our training set.

In particular, we train a multi-level or hierarchical Bayesian model combining information from different CCs and stellar effective temperature, metallicity, and gravity to estimate ages. There are two main advantages of using this technique in this context. The first advantage is that it naturally combines information from different linear regressions: as such, we do not need to choose one particular CC, and its related regression, over another. If different observations of CCs are available, this technique takes the information provided by all of them into account in the stellar age estimation. The second advantage is the proper and consistent treatment of uncertainties, ensuring reliable age uncertainty estimations.

\section{Training data sample}
\label{sec:data}
The data sample we used as the training sample is well described in DM19 and references therein. It consists of 1059 stars observed within the HARPS-GTO planet search program. These stars belong to a volume-limited sample around 70 pc of the Sun with very few stars at greater distances, ensuring that the relations we were going to determine are applicable to all these stars. The final spectra have a resolution of R ~115 000 and high S/N (45$\%$ of the spectra have 100 < S/N < 300, 40$\%$ of the spectra have S/N > 300, and the mean S/N is 380). Stellar parameters such as $T_{\rm eff}$, [Fe/H], and $\log g$ were derived using a special set of iron lines \citep[see][for details]{Elisa17}. The chemical abundances [X/Fe] were determined under local thermodynamic equilibrium (LTE) using the 2014 version of the code MOOG \citep{Sneden} and a grid of Kurucz ATLAS9 atmospheres \citep{Kurucz}.

Stellar masses and ages were obtained with the PARAM v1.3 tool using the PARSEC isochrones \citep{Bressan} and a Bayesian estimation method \citep{Silva06} together with the values for $T_{\rm eff}$ and [Fe/H] from \citet{Elisa17}, $V$ magnitudes from the main HIPPARCOS catalogue \citep{Perryman}, and parallaxes from the second Data Release (DR2) of Gaia \citep{Gaia16, Gaia18, Lindegren}, which are available for 1057 out of 1059 stars.

Not all the ages derived for these 1059 stars are reliable using this method. DM19 decided to define reliable age estimates as those with an age uncertainties smaller than 1.5 Ga. In this work, we also filtered out stars with age uncertainties smaller than 0.2 Ga, which we regard as being unrealistic for standard isochrone fitting, which can erroneously bias our final model. These cuts leave 328 out of 1059 stars for use in our studies. We refer to DM19 for details of the main characteristics of this subset. In summary, we worked with 244 thin-disk stars, 14 high-$\alpha$ metal-rich stars, 68 thick-disk stars, and 2 halo stars. These classifications were made following \citet{Vardan11,Vardan12}. These 328 stars also have a wide range in parameters $T_{\rm eff}$: 5010-6788 K (95$\%$ between 5271 and 6416 K), $\log g$: 3.73-4.71 dex (95$\%$ between 3.93 and 4.58 dex), and [Fe/H]: -1.15-0.55 dex (95$\%$ between -0.81 and 0.33 dex). In terms of uncertainties, e$T_{\rm eff}$ has an exponential distribution between 61-107 K (95$\%$ between 61 and 79 K), e$\log g$ is also distributed exponentially between 0.1-0.22 dex (95$\%$ between 0.1 and 0.12 dex), and 95\% of the\ e[Fe/H] has a 95$\%$  values lie in the range 0.04-0.05 dex; only a few values lie around 0.06 and 0.07 (maximum $\Delta$[Fe/H] of the sample).

\section{Inference technique: Hierarchical Bayesian model}
\label{sec:hbm}
\begin{figure*}
\begin{center}
\includegraphics[scale=0.3]{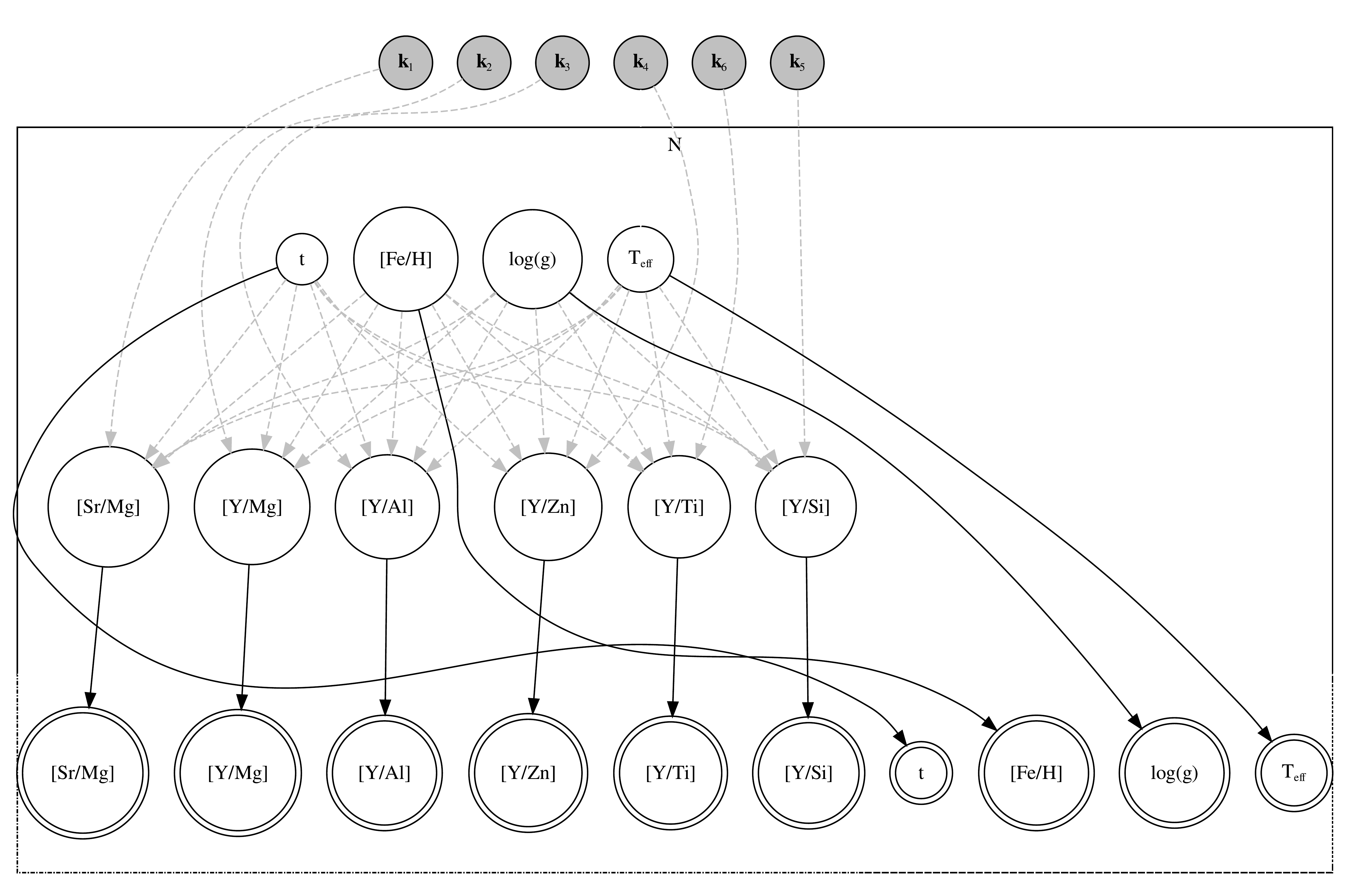}
\caption{Logical structure of the hierarchical Bayesian model we trained for this study. See text for details.}\label{fig:graph}
\end{center}
\end{figure*}

We defined a Bayesian hierarchical model that is graphically described in Fig.~\ref{fig:graph} and in the following paragraphs. Inspired by the multi-dimensional linear relations described in DM19 and given the indication in \citet{dotter} and \citet{gavel} that all of the potential predictive variables may carry a piece of physical information, we followed \cite{gelmanbda04} and included all available predictors with priors centred at zero. This is in practice equivalent to a regularisation that will only produce regression coefficient posteriors that are effectively different from zero if the data support them \citep[see][]{gelmanbda04}. In the top layer of the model, we have then the true values of the stellar physical parameters (hereafter stellar parameters) effective temperature $T_{\rm eff}$, iron abundance [Fe/H], (logarithm of the) surface gravity $\log g$, and age $t$. The true values of the abundance ratios used as CCs are (deterministically) modelled as a linear combination of these four stellar parameters,

\begin{equation}
    [R_i] = k_{i,0}+k_{i,1}\cdot t+k_{i,2}\cdot T_{\rm eff}+k_{i,3}\cdot{\rm [Fe/H]}+k_{i,4}\cdot\log g 
    \label{deterministic}
,\end{equation}

\noindent where $R_i$ is the i-th abundance ratio used as CC and $k_{i,j}$ is the $j$-th constant coefficient of the linear combination. The five coefficients $k_{i,j}$ of each linear combination are also model parameters.  

Finally, observables are set at the lowest level of the model and are defined as random variables normally distributed around the true values and with a standard deviation given by the measurement uncertainties described in Sect.~\ref{sec:data}. This is the most reasonable approximation possible for the distribution of these observables since we do not have their probabilistic distributions but their measurements and standard deviations.

We used the training set described in DM19 in order to infer the posterior distributions of the $k_{i,j}$ coefficients that were subsequently used to predict ages for other stars not in the training set. We refer to the first stage (inferring posterior distributions for the $k_{i,j}$ coefficients) as the training phase and to the second stage (applying the model to infer ages of stars not in the training set) as the prediction stage. 

Following DM19, we used five $\alpha$, odd- and even-Z element abundances (Mg, Ti, Al, Si, and Zn) on the one hand, and two s-process element abundances (Y and Sr) on the other to obtain the CCs. This allows the definition of ten ratios. Because we combined information of all of them, we must note that only six are linearly independent of the rest. We selected the five ratios that involve the Y abundance ([Y/Si], [Y/Mg], [Y/Ti], [Y/Zn], and [Y/Al]) and [Sr/Mg] as CCs because Y is usually easier to obtain than Sr, and its values are also usually more precise. In the following we refer to the vector of CCs thus defined for the $i$-th star as $\hat{\mathbf c}_i$ , where the circumflex denotes observed values. The CCs of our training set are affected by two sources of random noise. One source is those physical parameters not accounted for in our model, such as how well the ISM is mixed within the Galaxy or if the material is not well mixed \citep{Vardan15}. We refer to this as the intrinsic scatter. The second source is the measurement uncertainties. In our model  we assumed that the intrinsic scatter is much smaller than the measurement uncertainties and cannot be constrained from the observations. Hence, only the latter was included explicitly. 

The model then contains 328$\times$4 parameters that correspond to the true values of the stellar parameters, plus 6$\times$5 parameters that correspond to the linear combination coefficients for each CC. We denote the vector of true values of the stellar parameters of the $i$-th star as ${\bm \theta_i}$. As before, we use the circumflex to denote observed values. The likelihood function is then defined as

\begin{align}
\begin{split}
\mathcal{L} = & \prod_{i=1}^{328} p(\hat{\mathbf c}_i|{\mathbf c}_i)\cdot p(\hat{\bm \theta_i}|{\bm \theta_i}) = \prod_{i=1}^{328} p(\hat{\mathbf c}_i \mid {\bm \theta_i,{\bm K}})\cdot p(\hat{\bm \theta_i}|{\bm \theta_i}),\\
\end{split}
\end{align}

\noindent where we denote with ${\mathbf K}$ the set of six vectors of coefficients ${\mathbf k}_i$. The posterior probability distribution of the model parameters is then obtained applying Bayes' rule,

\begin{equation}
    p({\bm K},{\bm \Theta}|\hat{\mathbf c},\hat{\bm \Theta}) \propto {\mathcal L}\cdot \pi({\bm \Theta})\cdot \pi({\bm K})
\label{eq:post}
,\end{equation}

\noindent where we use the notation $\pi(\cdot)$ instead of $p(\cdot)$ to denote prior probability distributions, and we use ${\mathbf c}$ and ${\bm \Theta}$ to denote the set of 328 CC values and stellar parameters, respectively. 

We defined multivariate normal priors for each of the ${\mathbf k}_i$ vectors centred at the values of a maximum likelihood linear fit to the data, ${\mathbf K}_{ML}$. The covariance matrix ${\bm \Sigma}_K$ is a modification of the maximum likelihood fit covariance matrix, whereby the original diagonal is scaled (multiplied) by ten to make the prior significantly less informative. Hence, 

\begin{equation}
        \pi({\bm K}) = \prod_{j=1}^{6} {\mathcal N}({\mathbf k_j}|{\mathbf k}_{j,ML},{\bm \Sigma}_{j})
,\end{equation}

\noindent where $\mathcal{N}(\cdot \mid {\bm \mu},{\bm \Sigma})$ denotes the multi-variate Gaussian probability distribution centred at ${\bm \mu}$ and with covariance matrix ${\bm \Sigma}$, and the subscript $ML$ denotes the maximum likelihood solution.

We used a non-informative multivariate Gaussian prior for $T_{\rm eff}$, [Fe/H] , and $t$ to account for the known correlations existing between the three, and defined an independent prior for the surface gravity, $\log g$. The reason for this election can be found in the appendix. Correlations between CCs and stellar parameters are not taken into account because in our HBM, each CC follows a deterministic relation with the independent variables, not a probabilistic one. The multivariate Gaussian was centred at the mean of the observed values and the covariance matrix was added as an additional model parameter. We decomposed the covariance matrix into a scale and a correlation matrix \citep[see e.g.][]{GellmanHill},

\begin{equation}
{\bm \Sigma} = {\bm D}\cdot\Omega\cdot{\bm D}
,\end{equation}

\noindent where ${\mathbf D}$ is a diagonal matrix with a scale for each stellar parameter, and ${\bm \Omega}$ is the correlation matrix. We defined an LKJ prior \citep{LEWANDOWSKI20091989} with shape parameter $\eta = 1$ for the correlation matrix, representing the equivalent to a uniform distribution on correlations. Finally, we defined a Cauchy prior on the scales centred at 0 and $\gamma=5$. We omit these so-called hyperparameters in Fig. \ref{fig:graph} for the sake of clarity. This model is hierarchical in the sense that the prior on the stellar parameters is learnt from the data. Hence, there are two levels in the model: one level for the model parameters (coefficients and true stellar parameters), and one level for the hyperparameters (the covariance matrix of $T_{\rm eff}$, [Fe/H], and $t$ multivariate normal, having a LKJ prior).

In practice, we used {\it Stan} \citep{Stan} and the NUTS version of the Hamiltonian Monte Carlo (HMC) sampler in order to obtain samples of the posterior distribution defined in Eq. \ref{eq:post}. A more detailed analysis of the model and the reasons for the different choices we made can be found in the appendix.

The same model as described in Fig. \ref{fig:graph} can be used to predict the ages of stars not in the training phase. In this case, we are interested in the posterior probability density of the age given a set of observations of the remaining physical parameters and CCs. In the prediction, we did not aim to infer the distribution of the linear combination coefficients and instead used the distribution obtained in the training stage. We also used the posterior distribution of the covariance matrix $\Sigma$ for the prior of the stellar parameters ${\bm \theta}$. Let ${\bm \theta'}$ be the set of true stellar parameters excluding the age (i.e. $T_{\rm eff}$, $\log g$ , and $[Fe/H]$) and ${\hat{\bm \theta}'}$ the vector of observed values. The posterior distribution of the age can be obtained from the full posterior by marginalising out the uninteresting parameters (${\bm \theta}'$ and the coefficients ${\bm K}$),

\begin{align}
 p(t\mid \hat{\bm c},\hat{\bm \theta}') = &
\int p(t, {\bm \theta}', {\bm K} \mid \hat{\bm c},\hat{\bm \theta}')\cdot{\rm d}T_{\rm eff}\cdot{\rm d}\log g\cdot{\rm d}[Fe/H]\cdot{\rm d}{\bm K},
\label{eq:marg}
 \end{align}

\noindent and using Bayes' rule,

\begin{align}
p(t, {\bm \theta}', {\bm K} \mid \hat{\bm c},\hat{\bm \theta}') \propto p(\hat{\bm c},\hat{\bm \theta}'\mid t, {\bm \theta}', {\bm K})\cdot \pi(t,{\bm \theta}')\cdot\pi({\bm K}).
\label{eq:lik}
 \end{align}

We approximated the integral in Eq. \ref{eq:marg} with a sum of terms evaluated at the posterior samples of ${\bm K}$ and $\Sigma$ obtained during the training stage. As stated above, we used the posterior draws for ${\bm K}$ as the prior $\pi({\bm K})$ in the prediction stage and the posterior samples of $\Sigma$ to define the prior for ${\bm \theta} = (t,{\bm \theta}'),$ and we use these samples to evaluate the likelihood term $p(\hat{\bm c},\hat{\bm \theta}'\mid t, {\bm \theta}', {\bm K})$. In Eq. \ref{eq:lik} the marginalisation over $\hat{t}$ is implicit. Finally, for reasons that will be clearly understood in Sec. \ref{outliers}, for each star we performed ten age estimates. The final result we used is the mean value of these ten realisations.

Our inferences are slightly dependent on priors. They are multivariate normal priors centred at the values of a maximum likelihood linear fit to the data, that is, they are dependent on our training set. Therefore, any prediction obtained with our model is not reliable for stars that are not represented in our training set, such as M stars, giants, and sub-giants.

\section{Testing samples}
\label{sec:testing}

To test the HBM constructed with the training sample described in Section \ref{sec:data}, we gathered four complementary testing samples comprised of stars not belonging to the training set, as described below.

\begin{itemize}
    \item {\it Twenty-three stars with 'reliable' ages}, twenty of them dated using asteroseismology, two belonging to the cluster M67, and the Sun. The main characteristics of these stars are shown in Table \ref{testing_stars}. Here we can see the effective temperature, surface gravity, [Fe/H], age, and the relative abundances of Mg, Al, Si, Zn, Ti, Sr, and Y with respect to iron, and their respective uncertainties. Not all the chemical abundances are known for all the stars, which makes our test more realistic. To treat the Sun as a standard field star from this testing sample, its uncertainties were deliberately inflated to be similar to those of the remaining stars. These data were obtained as follows:

    \begin{itemize}
        \item[] The general characteristics of the Sun were taken from \cite{Prsa}. For the abundances, we analysed the Sun as a star using a Vesta spectrum and the line-list presented in \citet{Sousa07,Sousa08}. This line-list was created by calibrating the $\log g$ of lines to obtain an adopted reference solar Fe abundance of 7.47 dex as measured in the solar ATLAS spectrum. Because we used the Vesta spectra from HARPS as Sun, the iron is slightly different, as is [Fe/H]=-0.02 using this line-list. Nevertheless, since we used the differences between these abundances, this zero-point is irrelevant. That is, for the Sun, all the CCs are equal to zero. The uncertainties we used are quite conservative and are representative of the typical uncertainties for this stellar type.
        \item[] The characteristics of the two stars in the open cluster M67 come from \citet{Liu}. These stars are two solar twins in an open cluster with an age similar to that of the Sun. The age of the cluster was taken from \citet{Yadav}.
        \item[] The characteristics of the eight KIC stars and 16 Cyg A and B were obtained from \citet{Nissen17} and \citet{Morel20}. Ages for all stars come from asteroseismology, and their abundances were obtained in a special campaign for Kepler legacy stars with the HARPS-N\footnote{High Accuracy Radial velocity Planet Searcher - North} spectrograph at the TNG\footnote{Telescopio Nazionale Galileo Galilei} in \citet{Nissen17} and with the \'echelle fibre-fed SOPHIE spectrograph installed at the 1.93 m telescope of the Observatoire de Haute Provence (OHP, France) in the case of \citet{Morel20}. Because \citet{Nissen17} did not report any uncertainty for $\log g$, we imposed a standard and conservative uncertainty of 0.05 dex. Ages in the case of \citet{Nissen17} were taken from \citet{victor17}, where six different stellar structure and evolution codes and fitting algorithms were used. They re-analysed a few stars using ASTFIT \citep{jcd08a, jcd08b} and BASTA \citep{victor15}. On the other hand, \citet{Morel20} used two different studies for stellar dating, again \citet{victor17} on the one hand, and \citet{orlagh17}, where the AMP algorithm \citep{travis09} was used, on the other hand. They finally used the main of these two estimates because no clear differences were found in general. For more details about the age determination performed in these works and the treatment of the discrepancies, we refer to these papers.
    \end{itemize}
    
    \item {\it Seventy-nine stars from \citet{Spina18}}. They analysed  high-resolution HARPS spectra with a high signal-to-noise ratio of 79 solar twin stars in order to study the formation and evolution of the Galactic disk through the chemical composition of these stars. They provided atmospheric parameters, ages, and chemical abundances of 12 neutron-capture elements (Sr, Y, Zr, Ba, La, Ce, Pr, Nd, Sm, Eu, Gd, and Dy). The remaining chemical abundances were presented in \citet{Bedell18}. In this case, stellar ages were obtained by fitting evolutionary models (isochrones) to classic stellar parameters using the $q^2$ code, as shown in \cite{Ramireza, Ramirezb}.

    \item {\it Ten Gaia benchmarks stars}. We selected Gaia benchmark stars with $T_{\rm eff}$, $\log g$, and [Fe/H] within the range of values covered by our training sample. This means a total of ten stars. The stellar parameters were taken from the Gaia benchmark papers \cite{Heiter15} and \cite{Jofre14}. The chemical abundances were derived in this work by applying the same methodology as in our previous works \citep{Elisa17, Vardan12} to very high quality ESPRESSO spectra obtained by \citet{Vardan20} and using the stellar parameters mentioned above. Ages were obtained from \citet{Sahlholdt}. In that work, the authors determined ages by balancing all the previous determinations in the literature and different estimations of their own from different isochrones. Therefore, every individual determination is a heterogeneous inference from many different techniques (isochrone fitting, gyrochronology, activity, and/or asteroseismology), and the techniques used are different from one star to the next. The main physical characteristics of these stars can be found in Table \ref{gaia_stars}.

    \item {\it One hundred and three stars in open clusters}. \citet{Casamiquela} studied the physical characteristics and abundances of 93 stars belonging to Hyades, Praesepe (NGC 2632), and Ruprecht 147, with ages of about 0.8 Ga for Hyades, in the range [0.7, 0.83] Ga for NGC 2632, and in the range [2, 2.5] Ga for Ruprecht 147, according to Table 1 of \citet{Casamiquela}. The spectra were collected from many different instruments such as UVES at VLT\footnote{UV-visual echelle spectrograph, Very Large Telescope}, FEROS at MPG\footnote{Fiberfed Extended Range Optical Spectrograph, Max Planck Gesellschaft}, HARPS, HARPS-N, FIES at NOT\footnote{FIbre-fed Echelle Spectrograph, Nordic Optical Telescope}, ESPaDOnS at CFHT\footnote{Echelle SpectroPolarimetric Device for the Observation of Stars, Canada France Hawaii Telescope}, NARVAL at TBL\footnote{Telescope Bernard Lyot}, and ELODIE at OHP. For the abundances, they  used stars as reference that were similar in terms of their $T_{\rm eff}$ and [Fe/H] to those of the Hyades cluster. In adition, we included stars from \citet{Blanco}. They studied a total of 207 stars belonging to 34 open clusters with NARVAL, with ~300-1100 nm and an average resolution of ~80000; with HARPS, with ~378-691 nm and a gap between chips that affects the region from 530 to 533 nm and a resolution of ~115000; and with UVES, with ~476-683 nm and a small gap between 580 and 582 nm and a minimum resolution of ~47000. Unfortunately, the vast majority of stars observed from these clusters were giant stars, and only 93 from \citet{Casamiquela} and 10 in the cluster IC4651 from \citet{Blanco} can be analysed using our model mainly because of the $\log g$ range that is covered by our training sample. We also discarded clusters in which the number of remaining stars after filtering was statistically not significant. The ages of the clusters were taken from \citet{Dias}
\end{itemize}

\begin{sidewaystable*}
        \caption{Physical characteristics of the first set of testing stars 23 stars with 'reliable' ages. The symbol [X] represents the element X abundance with respect to iron ([X/Fe]), and "e[X]" represents the uncertainty of this measurement.}
        \label{testing_stars}
        \centering
        \resizebox{\columnwidth}{!}{
                \begin{tabular}{lp{0.02\linewidth}p{0.015\linewidth}p{0.02\linewidth}p{0.02\linewidth}rp{0.03\linewidth}rp{0.02\linewidth}rp{0.02\linewidth}rp{0.02\linewidth}rp{0.02\linewidth}rp{0.02\linewidth}rp{0.02\linewidth}rp{0.02\linewidth}p{0.02\linewidth}p{0.02\linewidth}}
                        \hline
                        star & $T_{\rm eff}$ & e$T_{\rm eff}$ & $\log g$ & e$\log g$ & [Fe/H] & e[Fe/H] & [Mg] & e[Mg] & [Al] & e[Al] & [Si] & e[Si] & [TiI] & e[TiI] & [Zn] & e[Zn] & [Sr] & e[Sr] & [Y] & e[Y] & Age & eAge\\
                        \hline
                        Sun & 5772 & 50 & 4.44 & 0.05 & -0.02 & 0.05 & -0.02 & 0.04 & -0.02 & 0.04 & -0.02 & 0.04& -0.02 & 0.04 & -0.02 & 0.04 & -0.02 & 0.04 & -0.02 & 0.04 & 4.60 & 0.50\\ 
                        M67 1194 & 5786 & 13 & 4.46 & 0.02 & -0.005 & 0.010 & -0.003 & 0.011 & 0.01 & 0.02 & -0.001 & 0.013& 0.01 & 0.01 & -0.03 & 0.02 & -0.01 & 0.02 & 0.02 & 0.02 & 4.15 & 0.65\\ 
                        M67 1315 & 5933 & 23 & 4.47 & 0.05 & -0.061 & 0.014 & -0.01 & 0.03 & 0.04 & 0.02 & -0.016 & 0.014& 0.04 & 0.02 & -0.03 & 0.02 & 0.07 & 0.02 & 0.01 & 0.02 & 4.15 & 0.65\\ 
                        KIC3427720$^{(a)}$ & 6074 & 20 & 4.39 & 0.05 & -0.03 & 0.02 & -0.01 & 0.01 & -0.03 & 0.02 & 0.01 & 0.01& 0.00 & 0.01 & -0.05 & 0.02 &  &  & -0.01 & 0.02 & 2.40 & 0.30\\ 
                        KIC3656476$^{(b)}$ & 5680 & 15 & 4.233 & 0.003 & 0.27 & 0.03 & -0.01 & 0.03 & 0.05 & 0.03 & 0.03 & 0.03 & -0.01 & 0.04 & 0.01 & 0.10 & -0.14 & 0.06 & -0.09 & 0.04 & 8.70 & 0.30 \\ 
                        KIC5184732$^{(b)}$ & 5855 & 24 & 4.266 & 0.006 & 0.41 & 0.04 & 0.05 & 0.08 & 0.03 & 0.05 & 0.04 & 0.04 & -0.01 & 0.05 & 0.03 & 0.13 & -0.04 & 0.07 & -0.06 & 0.05 & 4.49 & 0.45 \\ 
                        KIC6106415$^{(b)}$ & 6035 & 19 & 4.296 & 0.002 & -0.05 & 0.03 & -0.01 & 0.04 & 0.00 & 0.03 & 0.00 & 0.04 & 0.00 & 0.05 & -0.07 & 0.05 & -0.03 & 0.06 & -0.01 & 0.03 & 4.65 & 0.18 \\ 
                        KIC6225718$^{(a)}$ & 6287 & 18 & 4.32 & 0.05 & -0.11 & 0.01 & -0.02 & 0.01 & -0.03 & 0.01 & 0.007 & 0.009 & 0.02 & 0.01 & -0.06 & 0.02 &  &  & 0.03 & 0.01 & 2.60 & 0.40\\ 
                        KIC6603624$^{(b)}$ & 5625 & 20 & 4.325 & 0.004 & 0.27 & 0.04 & 0.02 & 0.06 & 0.07 & 0.04 & 0.02 & 0.04 & 0.00 & 0.06 & 0.08 & 0.09 & -0.11 & 0.07 & -0.04 & 0.04 & 8.50 & 0.46 \\ 
                        KIC7871531$^{(b)}$ & 5510 & 15 & 4.477 & 0.005 & -0.19 & 0.03 & 0.03 & 0.05 & 0.08 & 0.03 & -0.03 & 0.03 & 0.12 & 0.05 & 0.01 & 0.07 & -0.09 & 0.06 & -0.10 & 0.04 & 9.10 & 0.33 \\ 
                        KIC7940546$^{(a)}$ & 6326 & 15 & 4.00 & 0.05 & -0.13 & 0.01 & -0.04 & 0.01 & -0.03 & 0.01 & 0.027 & 0.008 & 0.020 & 0.009 & -0.05 & 0.02 &  &  & 0.02 & 0.01 & 2.40 & 0.30\\ 
                        KIC7970740$^{(b)}$ & 5365 & 15 & 4.545 & 0.003 & -0.47 & 0.03 & 0.20 & 0.06 & 0.23 & 0.04 & 0.09 & 0.04 & 0.28 & 0.07 & 0.14 & 0.09 & -0.05 & 0.07 & 0.01 & 0.03 & 10.68 & 0.59 \\
                        KIC8006161$^{(b)}$ & 5415 & 23 & 4.497 & 0.002 & 0.35 & 0.08 & 0.05 & 0.06 & 0.11 & 0.07 & 0.04 & 0.08 & 0.05 & 0.07 & 0.20 & 0.19 & 0.00 & 0.17 & -0.02 & 0.06 & 4.61 & 0.32 \\ 
                        KIC8694723$^{(b)}$ & 6295 & 56 & 4.112 & 0.004 & -0.41 & 0.06 & 0.03 & 0.06 & -0.07 & 0.08 & 0.00 & 0.07 & 0.04 & 0.08 & -0.06 & 0.07 & -0.02 & 0.08 & -0.03 & 0.07 & 4.67 & 0.31 \\ 
                        KIC8760414$^{(b)}$ & 5985 & 35 & 4.329 & 0.003 & -0.95 & 0.05 & 0.30 & 0.07 &  &  & 0.16 & 0.05 & 0.18 & 0.05 & 0.01 & 0.12 & 0.05 & 0.07 & 0.00 & 0.05 & 12.00 & 0.29 \\ 
                        KIC9139151$^{(a)}$ & 6115 & 22 & 4.38 & 0.05 & 0.10 & 0.02 & -0.07 & 0.02 & -0.06 & 0.02 & 0.00 & 0.01& -0.01 & 0.01 & -0.05 & 0.03 &  &  & 0.04 & 0.02 & 1.90 & 0.70\\ 
                        KIC9965715$^{(b)}$ & 6335 & 40 & 4.280 & 0.004 & -0.29 & 0.04 & 0.01 & 0.07 & -0.03 & 0.07 & 0.02 & 0.06 & 0.15 & 0.05 & -0.19 & 0.04 & 0.02 & 0.07 & 0.04 & 0.07 & 3.40 & 0.80 \\ 
                        KIC10162436$^{(a)}$ & 6261 & 27 & 3.98 & 0.05 & -0.07 & 0.02 & -0.04 & 0.02 & -0.05 & 0.02 & 0.01 & 0.01 & 0.02 & 0.02 & -0.03 & 0.03 &  &  & 0.03 & 0.02 & 2.50 & 0.40\\ 
                        KIC10644253$^{(a)}$ & 6102 & 22 & 4.40 & 0.05 & 0.13 & 0.02 & -0.04 & 0.02 & -0.03 & 0.02 & -0.02 & 0.01& -0.01 & 0.02 & -0.07 & 0.03 &  &  & 0.03 & 0.02 & 1.30 & 0.70\\ 
                        16 Cyg A$^{(a)}$ & 5816 & 10 & 4.29 & 0.05 & 0.093 & 0.007 & 0.03 & 0.01 & 0.035 & 0.008 & 0.01 & 0.02& 0.031 & 0.007 & 0.019 & 0.009 &  &  & -0.057 & 0.007 & 7.00 & 0.50\\ 
                        16 Cyg B$^{(a)}$ & 5763 & 10 & 4.36 & 0.05 & 0.062 & 0.007 & 0.03 & 0.01 & 0.043 & 0.008 & 0.01 & 0.02& 0.035 & 0.007 & 0.020 & 0.009 &  &  & -0.052 & 0.007 & 7.10 & 0.50\\ 
                        KIC12258514$^{(a)}$ & 6020 & 50 & 4.12 & 0.07 & 0.03 & 0.04 & 0.00 & 0.02 & 0.02 & 0.02 & 0.03 & 0.01& -0.01 & 0.01 & -0.01 & 0.02 &  &  & -0.05 & 0.02 & 4.50 & 0.80\\ 
                        KIC12317678$^{(b)}$ & 6550 & 112 & 4.06 & 0.01 & -0.19 & 0.12 & -0.01 & 0.08 & -0.13 & 0.14 & -0.07 & 0.13 & 0.14 & 0.11 & -0.12 & 0.15 & 0.13 & 0.09 & 0.25 & 0.28 & 2.35 & 0.40 \\
                        \hline
        \end{tabular}}
        \tablebib{(a)~ \citet{Nissen17}; (b)~ \citet{Morel20}.}
\end{sidewaystable*}

\begin{table*}
        \caption{Physical characteristics of the Gaia benchmark testing stars. The symbol [X]$_{\rm H}$ represents the element X abundance with respect to hydrogen ([X/H]), and "e[X]$_{\rm H}$" represents the uncertainty of this measurement. The stellar parameters were taken from the Gaia benchmark papers \cite{Heiter15} and \cite{Jofre14}. The chemical abundances were derived in this work. See text for details.}
        \label{gaia_stars}
        \centering
        \begin{tabular}{lrrrrrrrrrrrr}
                \hline
                star & $T_{\rm eff}$ & e$T_{\rm eff}$ & $\log$g & e$\log$g & [Fe/H] & e[Fe/H] & [Mg]$_{\rm H}$ & e[Mg]$_{\rm H}$ & [Al]$_{\rm H}$ & e[Al]$_{\rm H}$ & [Si]$_{\rm H}$ & e[Si]$_{\rm H}$\\
                \hline
                18 Sco & 5810 &  80 & 4.44 & 0.03 & 0.01 & 0.03 & 0.00 & 0.02 & -0.02 & 0.04 & 0.05 & 0.02\\
                HD22879 & 5868 &  89 & 4.27 & 0.04 & -0.88 & 0.05 & -0.49 & 0.04 & -0.65 & 0.04 & -0.55 & 0.04\\ 
                $\alpha$ Cen A & 5792 &  16 & 4.31 & 0.01 & 0.24 & 0.08 & 0.20 & 0.03 & 0.19 & 0.05 & 0.28 & 0.06\\
                $\alpha$ Cen B & 5231 &  20 & 4.53 & 0.03 & 0.22 & 0.10 & 0.19 & 0.04 & 0.12 & 0.01 & 0.29 & 0.04\\
                $\beta$ Hyi & 5873 &  45 & 3.98 & 0.02 & -0.07 & 0.06 & -0.04 & 0.01 & -0.05 & 0.01 & -0.03 & 0.02\\
                $\beta$ Vir & 6083 &  41 & 4.10 & 0.02 & 0.21 & 0.07 & 0.11 & 0.03 & 0.11 & 0.01 & 0.18 & 0.05\\ 
                $\epsilon$ Eri & 5076 &  30 & 4.61 & 0.03 & -0.10 & 0.06 & -0.17 & 0.06 & 0.01 & 0.04 & -0.04 & 0.05\\ 
                $\mu$ Ara & 5902 &  46 & 4.30 & 0.03 & 0.33 & 0.13 & 0.32 & 0.04 & 0.40 & 0.02 & 0.36 & 0.03\\
                Procyon & 6554 &  84 & 4.00 & 0.02 & -0.04 & 0.08 & 0.01 & 0.06 & -0.14 & 0.05 & 0.09 & 0.04\\
                $\tau$ Cet & 5414 &  21 & 4.49 & 0.02 & -0.50 & 0.03 & -0.28 & 0.05 & -0.23 & 0.01 & -0.37 & 0.04\\ 
                \hline
                \hline
                & [Ti]$_{\rm H}$ & e[Ti]$_{\rm H}$ & [Zn]$_{\rm H}$ & e[Zn]$_{\rm H}$ & [Sr]$_{\rm H}$ & e[Sr]$_{\rm H}$ & [Y]$_{\rm H}$ & e[Y]$_{\rm H}$ & Age & eAge\\ 
                \hline
                18 Sco & 0.04 & 0.02 & -0.02 & 0.03 & 0.04 & 0.08 & 0.04 & 0.02 &4.0 & 1.0\\ 
                HD22879 & -0.51 & 0.03 & -0.70 & 0.03 & -0.69 & 0.08 & -0.82 & 0.03 & 10.0 & 2.0\\ 
                $\alpha$ Cen A & 0.21 & 0.03 & 0.25 & 0.01 & 0.13 & 0.08 & 0.17 & 0.03 & 5.5 & 1.5\\ 
                $\alpha$ Cen B & 0.26 & 0.07 & 0.16 & 0.06 & 0.21 & 0.17 & 0.11 & 0.11 & 5.5 & 1.5\\ 
                $\beta$ Hyi & -0.04 & 0.03 & -0.02 & 0.03 & -0.14 & 0.08 & -0.09 & 0.06 & 6.0 & 1.0\\
                $\beta$ Vir & 0.09 & 0.02 & 0.16 & 0.03 & 0.07 & 0.08 & 0.18 & 0.08 & 3.0 & 1.0\\ 
                $\epsilon$ Eri & -0.05 & 0.05 & -0.18 & 0.07 & 0.06 & 0.17 & -0.03 & 0.11 & 0.65 & 0.25\\ 
                $\mu$ Ara & 0.41 & 0.04 & 0.34 & 0.03 & 0.30 & 0.08 & 0.26 & 0.04 & 6.0 & 2.0\\ 
                Procyon & 0.00 & 0.04 & 0.05 & 0.05 & 0.13 & 0.08 & 0.11 & 0.09 & 2.0 & 0.5\\
                $\tau$ Cet & -0.21 & 0.03 & -0.46 & 0.05 & -0.56 & 0.08 & -0.68 & 0.06 & 7.0 & 3.0\\ 
                \hline
        \end{tabular}
\end{table*}

For all these stars, except those from \citet{Casamiquela}, the different authors performed a differential analysis with respect to the Sun to obtain abundances. Therefore, the [X/H] values are all in the same scale. \citet{Casamiquela} used two approaches. First, they derived abundances with respect to the Sun in the same way as the other works, and these are the [X/H] values they provide. Later, however, they performed a differential analysis with respect to reference stars in the Hyades.

With these four testing samples, we analysed the performance of our HBM in estimating stellar ages. The first three testing sets (i.e., the 23 stars with 'reliable' ages, 79 stars from \citet{Spina18}, and 10 Gaia benchmarks stars) tested the performance of our model on individual stars (separated in terms of the techniques used for estimating their ages and their astrophysical interest). The testing sample using stellar clusters was used to assess the statistical performance of our model, that is, how it estimate ages for a large set of stars with the same age, but very different stellar parameters.

\begin{figure*}
\begin{tabular}{cc}
\includegraphics[width=0.45\linewidth]{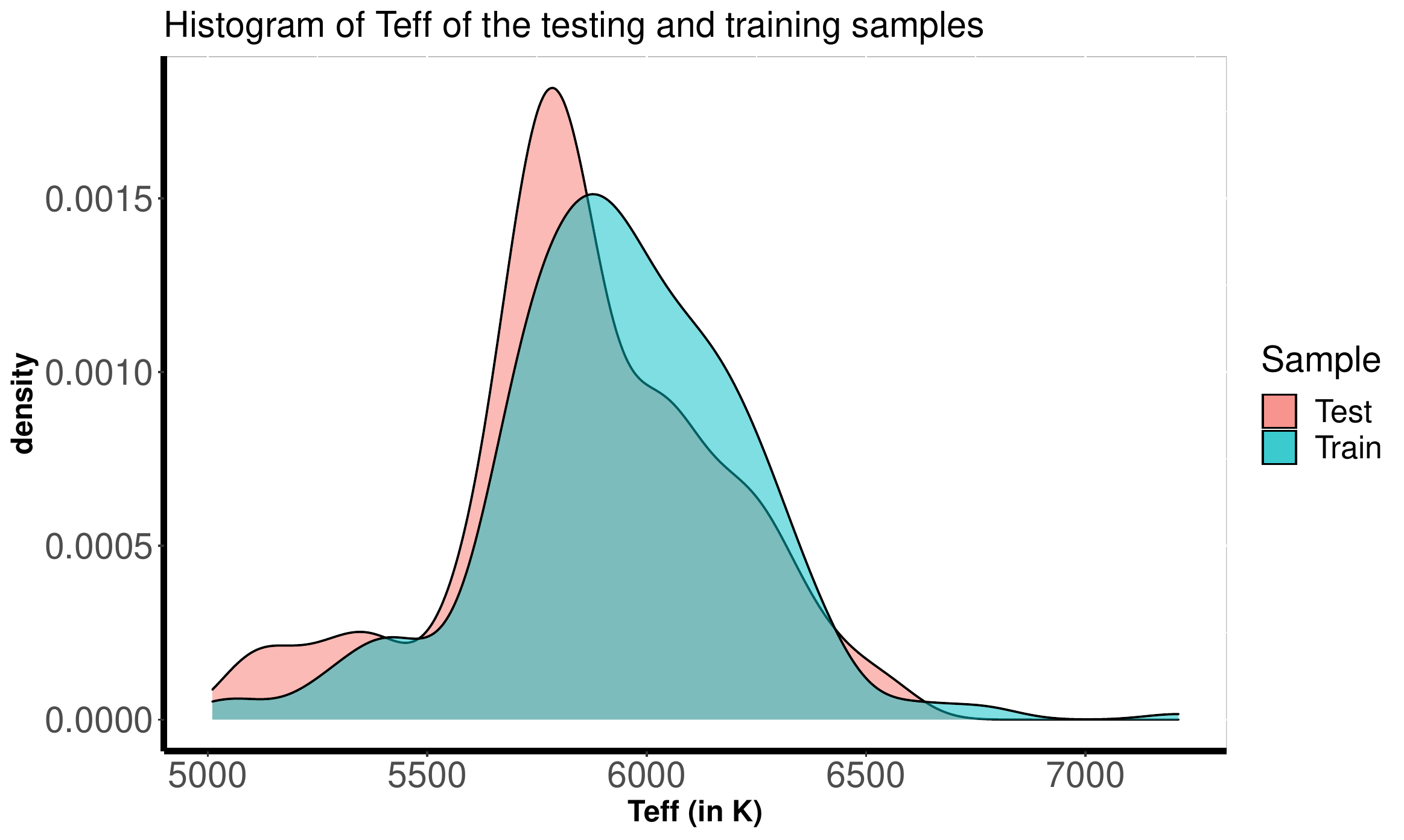}
     &  
\includegraphics[width=0.45\linewidth]{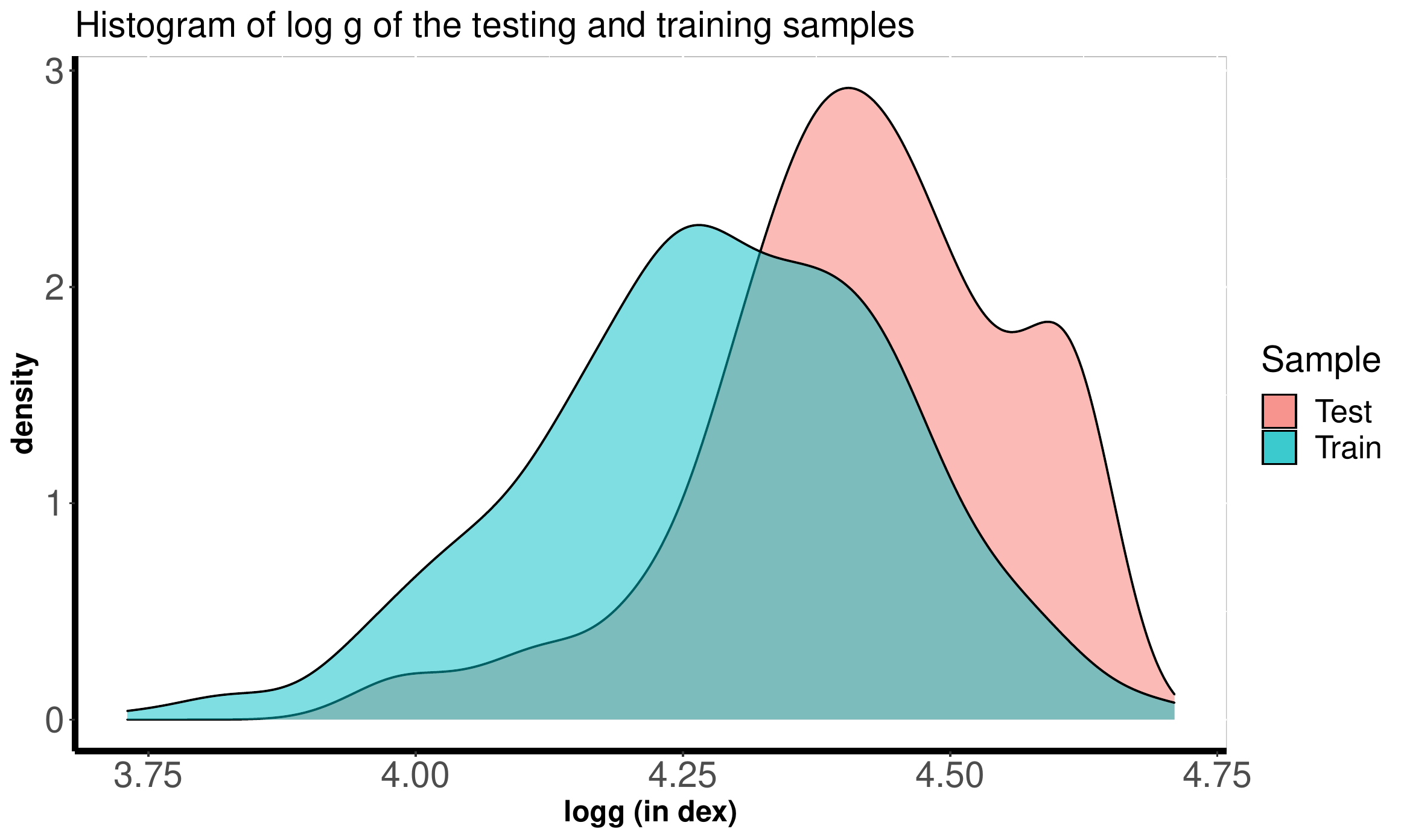}
     \\
\includegraphics[width=0.45\linewidth]{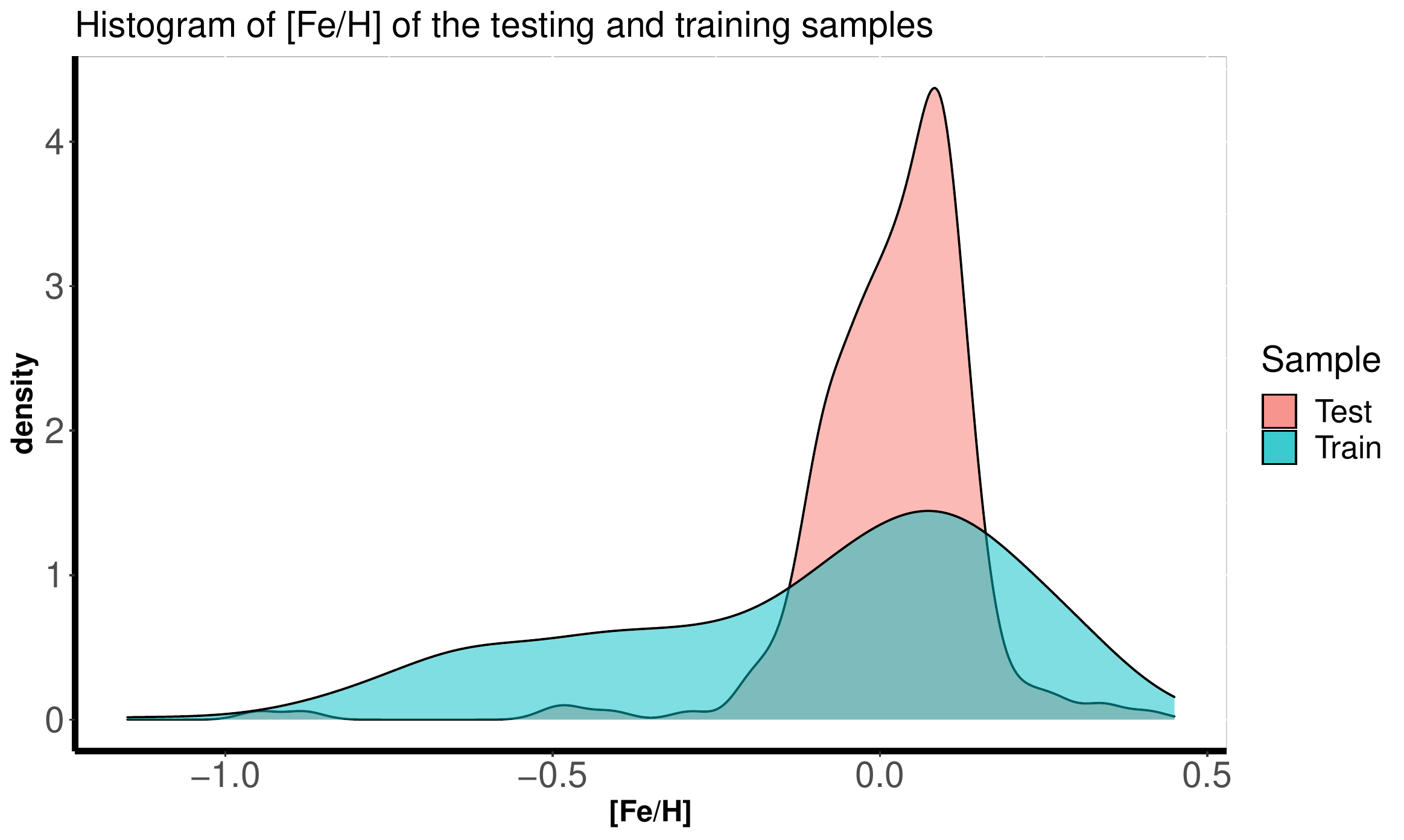}
     & 
\includegraphics[width=0.45\linewidth]{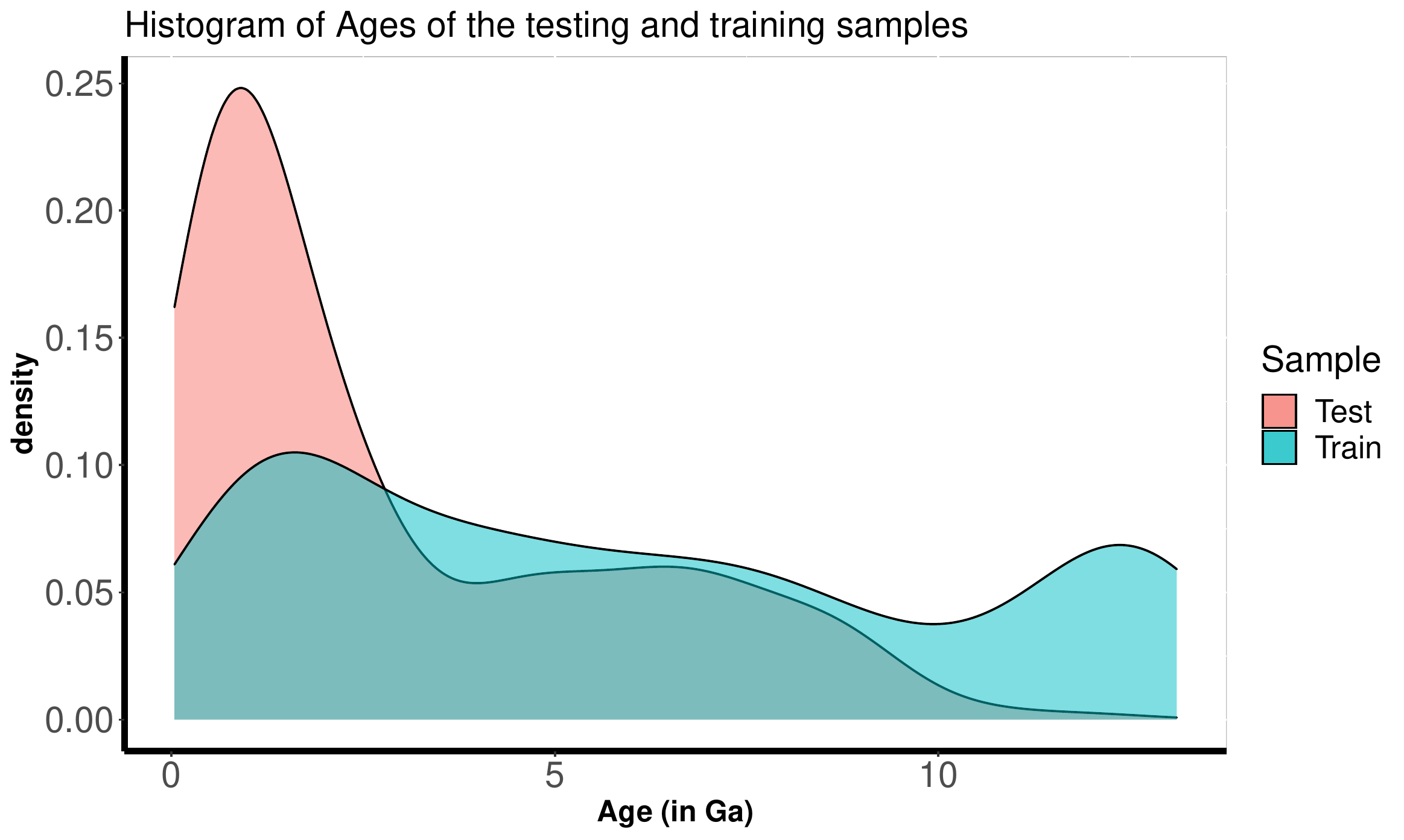}
\end{tabular}
\begin{center}
\caption{Histograms for $T_{\rm eff}$, $\log g$, [Fe/H], and age of the testing (red) and training samples (blue).}\label{histrograms}
\end{center}
\end{figure*}

Therefore, we gathered 215 stars for the testing sample covering many different situations in terms of input quality and age determination. To show how well this testing sample represents the training sample, Fig. \ref{histrograms} shows density distributions for $T_{\rm eff}$, $\log g$, [Fe/H], and age from the testing and training samples. This figure clearly shows that the testing sample properly covers the range of $T_{\rm eff}$ defined by the training sample. There is only a lack of stars with high temperatures, but this range is poorly described by the training sample. On the other hand, for $\log g$, there is a bias in the testing sample to large $\log g$. Nevertheless, almost all the $\log g$ space is tested. The situation is similar for the ages, where almost the entire range is covered by the testing sample, but with a clear overtesting of young ages, mainly because of the number of young cluster in our sample. The least covered variable is metallicity. The range covered by our training sample is far larger than that covered by the testing sample. The testing sample has a lack of low-metallicity stars. This bias must be corrected for in the future to offer a more consistent testing of the model, but it currently is a hard task to find testing stars in this context because stars with accurate abundances of the elements we need for this study and accurate ages are lacking. This last requirement is the most limiting one in general. This also limits the total number of testing stars we can use.

\section{Results and discussion}

In this section, we test the performance of the model described in section \ref{sec:hbm} that we obtained using the training set described in section \ref{sec:data}. These tests were made to compare the estimated ages in the literature for the testing stars described in section \ref{sec:testing} and the ages predicted by our model. Each subsection is devoted to each of the four testing sets presented in section \ref{sec:testing}. We also add a subsection presenting the combination of the results obtained for the three first testing sets (field stars). This is done to show the performance of our model for all the individual stars together, and to verify the consistency of our model compared with other dating methods in the literature. We finally analyse the outliers we found to understand their origin.

\subsection{Results for the 23 stars with 'reliable' ages}
\label{23_buenas}

\begin{figure}
\begin{center}
\includegraphics[width=\linewidth]{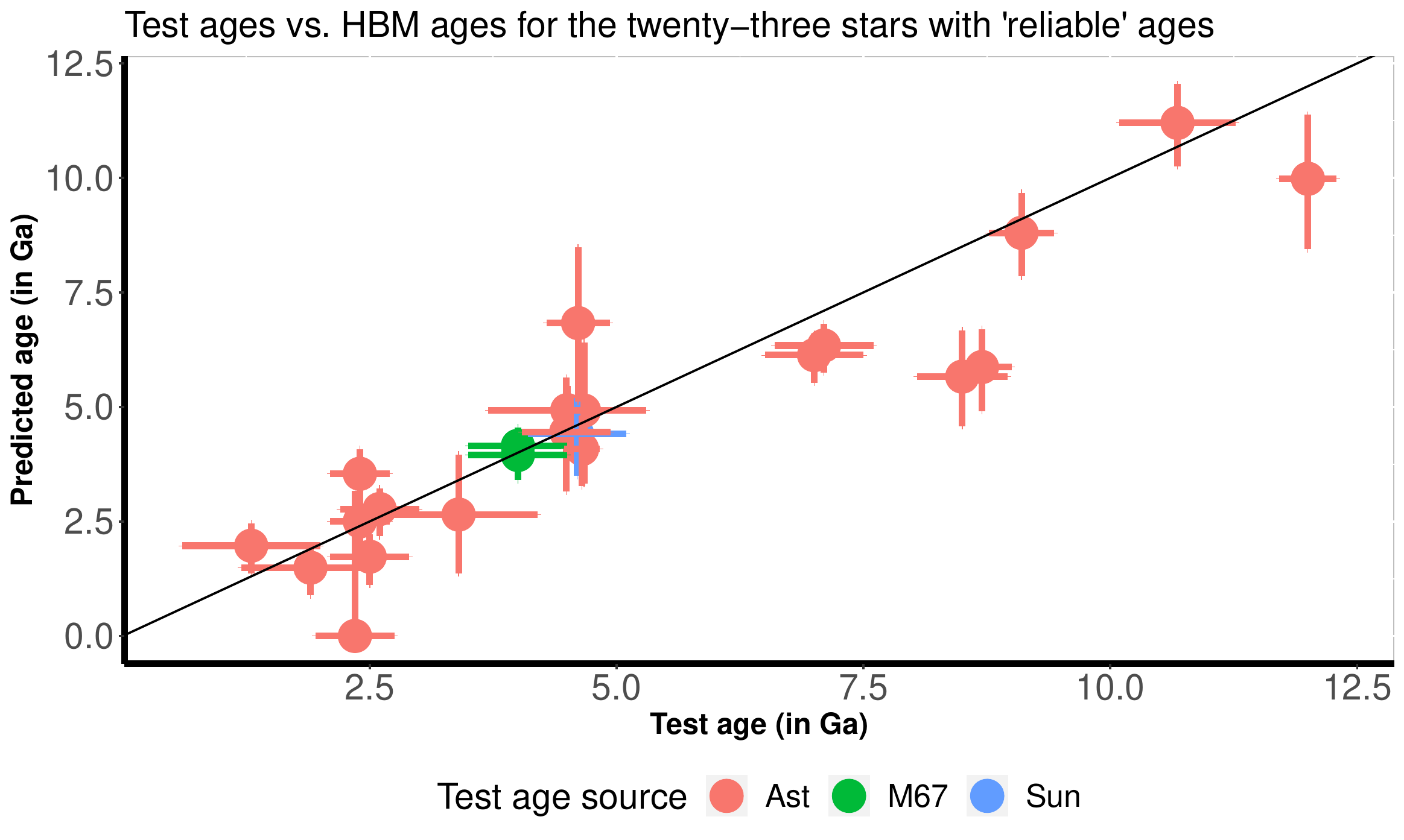}
\caption{Age predictions for the 23 stars with 'reliable' ages (test ages) using our HBM. The black line represents the one-to-one relation to guide the eye.}\label{23_pred}
\end{center}
\end{figure}

In Fig. \ref{23_pred} we show the comparison between the test ages that were obtained via asteroseismology, cluster membership, and so on with the ages predicted by the HBM. For the ages obtained in this work, uncertainties always represent the  $1\sigma$ dispersion. We evaluated the absolute differences, in the sense test minus estimated ages. We did not evaluate relative differences because uncertainties are not in general related to the age of the star. Therefore, the same uncertainty at two different ages may produce very different relative differences. For the complete set, we find a mean absolute difference (MAD) of 0.86 Ga. The mean difference (MD, a measure of the bias) is 0.19 Ga, that is, CCs slightly underestimate the age compared with asteroseismology, but this underestimation is within uncertainties. This agrees with the results of \citet{Morel20}, who found an MAD of 0.7 Ga with a bias towards younger ages estimated via CCs. A possible source for this bias is the fact that ages for the training sample were obtained using isochrones, as described in DM19, and with this test, we are evaluating the differences between asteroseismology and those isochrones plus CC predictors.

In addition, this figure shows some interesting features. The ages of the Sun and the two stars in M67 (blue and green dots in Fig. \ref{23_pred}) are predicted very well. Asteroseismic ages are also predicted with an MAD of 1 Ga. In general, for almost all the 23 stars, the ranges of test and predicted uncertainty overlap.

We identify three outliers with differences larger than 2 Ga and underctainties that do not overlap: KIC3656476, KIC6603624, and KIC8006161. These cases are discussed in Sect. \ref{outliers}.

\subsection{Results for the 79 stars from \citet{Spina18}}
\label{Spina}

\begin{figure}
\begin{center}
\includegraphics[width=\linewidth]{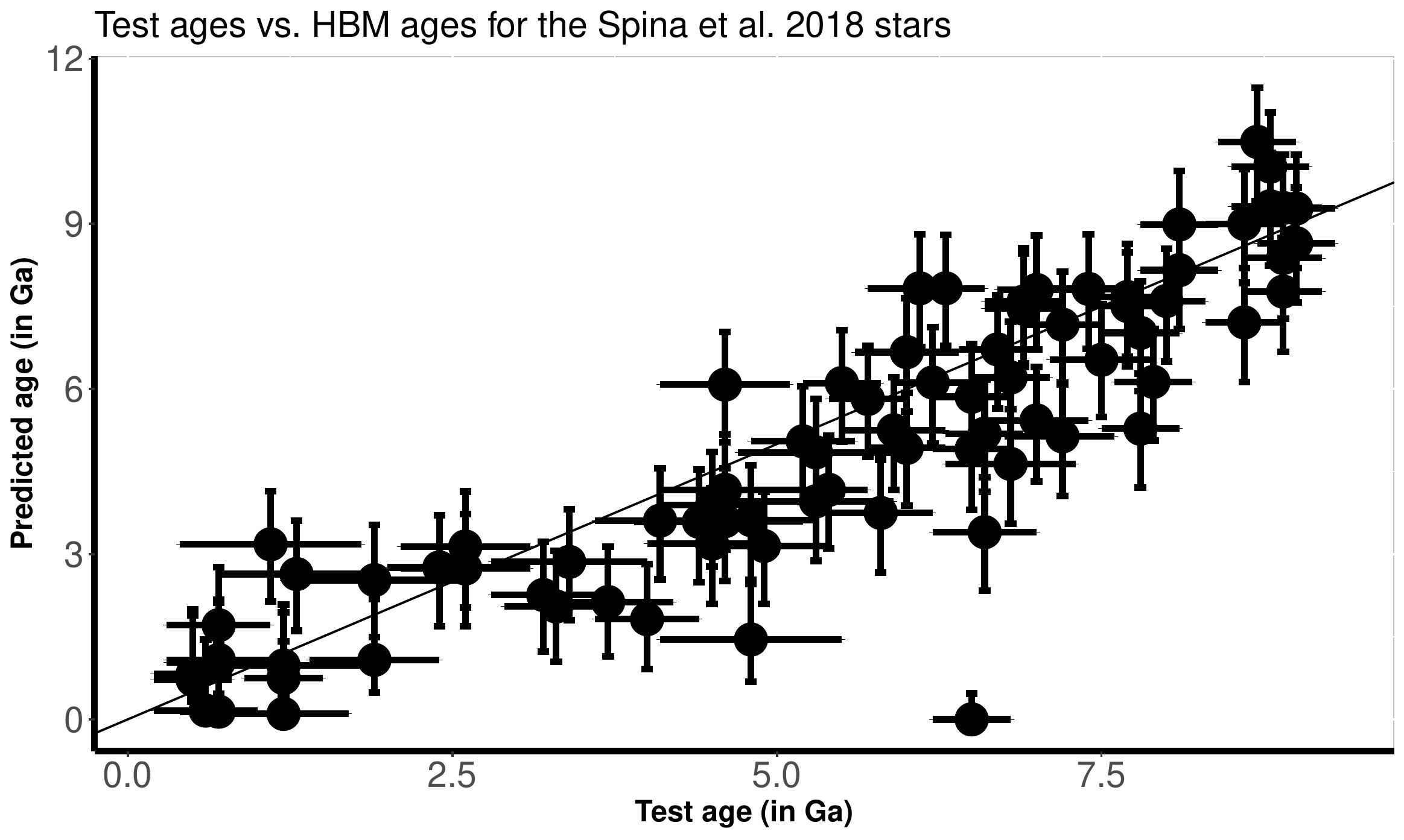}
\caption{Age predictions for the 79 stars from \citet{Spina18} (test ages) using our HBM. The black line represents the one-to-one relation to guide the eye.}\label{Spina_pred}
\end{center}
\end{figure}

In Fig. \ref{Spina_pred} we show the comparison between the ages estimated in \citet{Spina18} with the ages predicted by our HBM. The figure shows that the ages estimated by our model are similar to those provided by the authors using other isochrones. Nevertheless, we find a clear trend in this comparison that can be attributed to the different techniques used in \citet{Spina18} and DM19 to estimate ages from isochrones or to the different methods used by these authors to estimate abundances. Distinguishing these two options is beyond the scope of this work. On the other hand, we find a clear outlier. The star HIP64150 is also identified by \citet{Spina18} as a chemically anomalous star. These authors explained that this star belongs to a binary system in which the primary is orbited by a white dwarf. Therefore, its atmospheric abundances may be enhanced by pollution from an AGB companion \citep{Spina18}. Again, we evaluated the absolute differences without the outlier, finding an MAD of 0.93 Ga. and an MD of 0.38 Ga. That is, the bias is almost negligible within the errors, as for the previous testing set.

\begin{figure}
\begin{center}
\includegraphics[width=\linewidth]{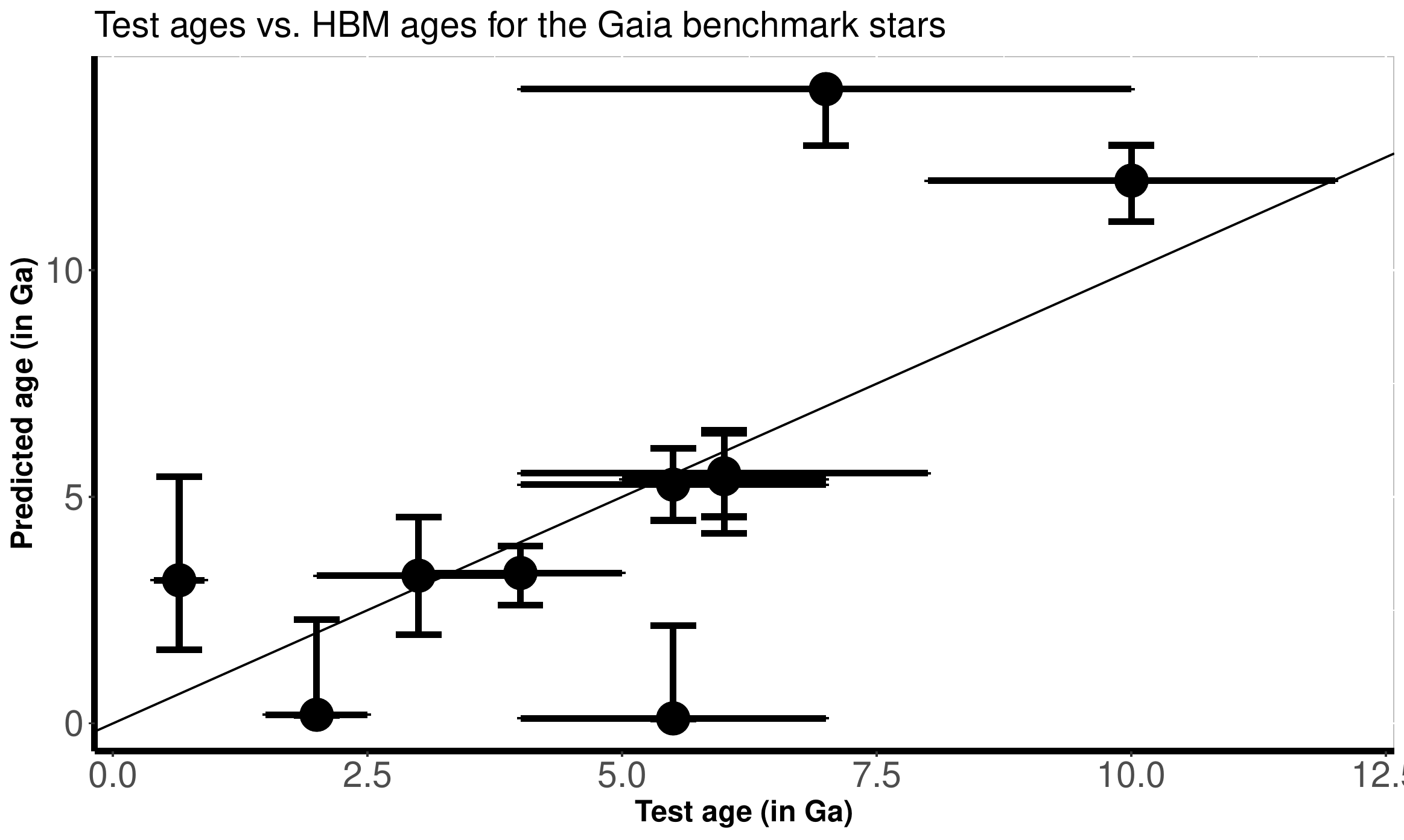}
\caption{Age predictions for ten Gaia benchmark stars (test ages are taken from \citet{Sahlholdt}) using our HBM. The black line represents the one-to-one relation to guide the eye.}\label{Gaia_pred}
\end{center}
\end{figure}

\subsection{Gaia benchmark stars}
\label{Gaia}

In the case of the Gaia benchmark stars, \citet{Sahlholdt} and references therein made a comprehensive study of estimates coming from many stellar dating techniques, such as asteroseismology, gyrochronology, stellar activity, and isochrone fitting, to name the most frequently used methods, and many sources in the literature, finally giving an age summary of all these estimates. Therefore, we can place our estimates in this context for a better understanding of our results compared with other techniques in the literature.

In Fig. \ref{Gaia_pred} we show the comparison between the ages estimated in \citet{Sahlholdt} with the ages predicted by our HBM. Here we find that for seven of the stars, the ages overlap (taking the uncertainties into account). There are three outliers ($\tau$ Cet, $\alpha$ Cen B, and $\epsilon$ Eri). These cases are analysed in Sect. \ref{outliers}.

Ignoring these three outliers, the remaining set has an MAD of 0.86 Ga and an MD of 0.22 Ga. That is, the bias is much lower than the uncertainty, and the MAD is of the order of or lower than 1 Ga, as was the case for the previous tests discussed above.

\begin{figure}
\begin{center}
\includegraphics[width=\linewidth]{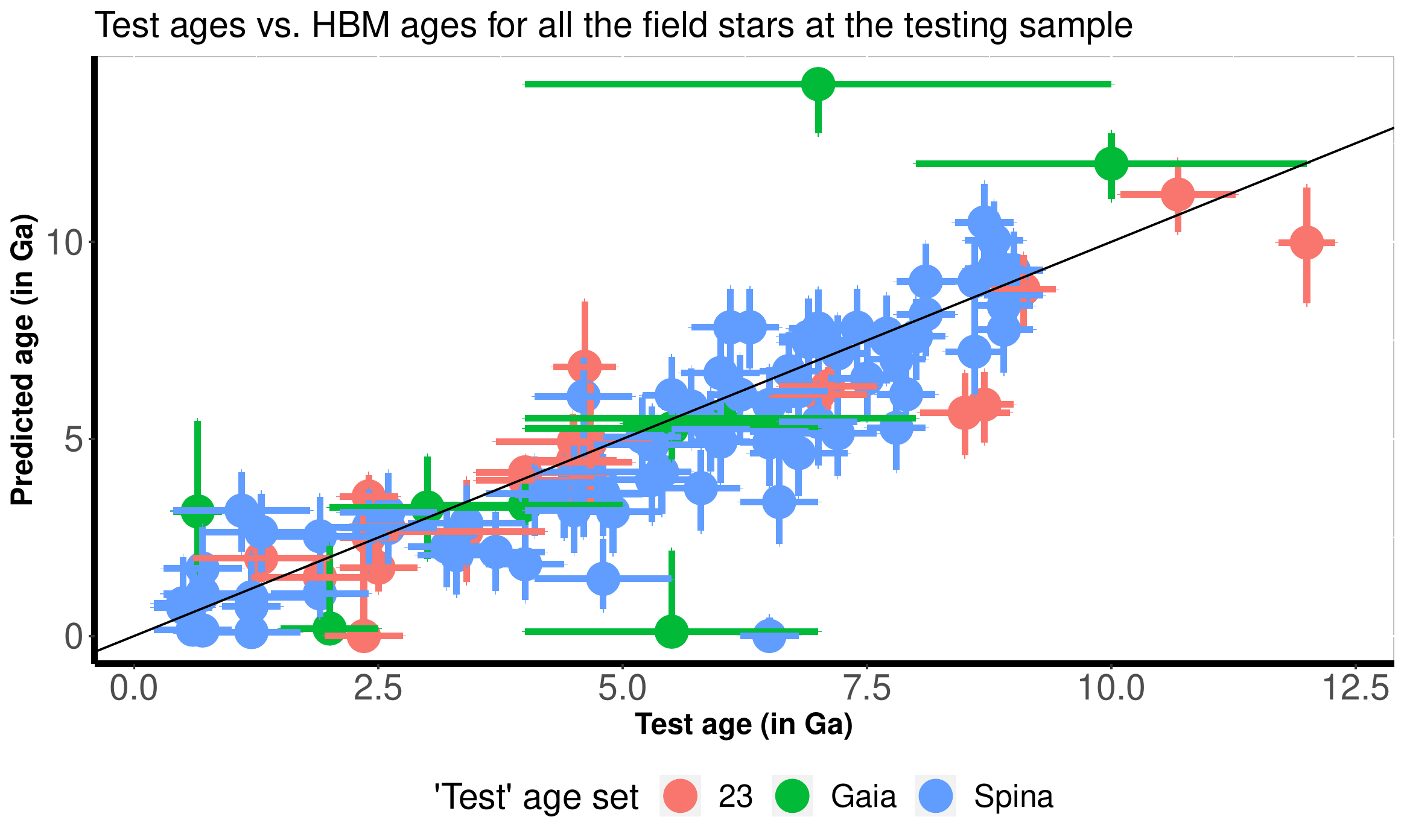}
\caption{Age predictions for the stars studied in Sects. \ref{23_buenas}, \ref{Spina}, and \ref{Gaia} using our HBM.}\label{all_tog}
\end{center}
\end{figure}

\subsection{All the field stars together}
\label{combination}

Finally, if we consolidate all the field stars studied in Sects. \ref{23_buenas}, \ref{Spina}, and \ref{Gaia} in a single plot, we obtain Fig. \ref{all_tog}. In addition to the already noted outliers, the results are quite stable, despite the fact that the age estimation techniques used for this testing sample are very heterogeneous. The trend found in the comparison with \citet{Spina18} is somewhat mitigated in the global dispersion, even when it is the subsample with the largest population. That is, it is not possible to distinguish the original dating technique in Fig. 6. This shows that this heterogeneity in estimation techniques for the testing stars has a negligible impact on the final results. Removing the Gaia benchmark outliers and the \citet{Spina18} single outlier, that is, those that can be clearly identified by comparison with another methods, we obtain an MAD of 0.93 Ga, and an MD of 0.35 Ga in this joint case. These values can be regarded as the performance of our HBM for estimating ages of field stars taking our testing sample into account.

\subsection{Stars in different open clusters}
\label{Clusters}

Dating using CCs is based on statistics. The idea is to analyse many stars and find a number of correlations between CCs, stellar parameters, and stellar age. One of the best ways of testing the statistical nature of the technique is to have many stars with many different physical parameters but the same age. Then the intrinsic dispersion of the method becomes visible.

In this section we propose using open clusters for this test. DM19 showed that the training sample has a clear dispersion and also outliers. Therefore, we can expect a similar statistical behaviour when stellar ages are estimated based on this training sample.

\begin{figure}
\begin{center}
\includegraphics[width=\linewidth]{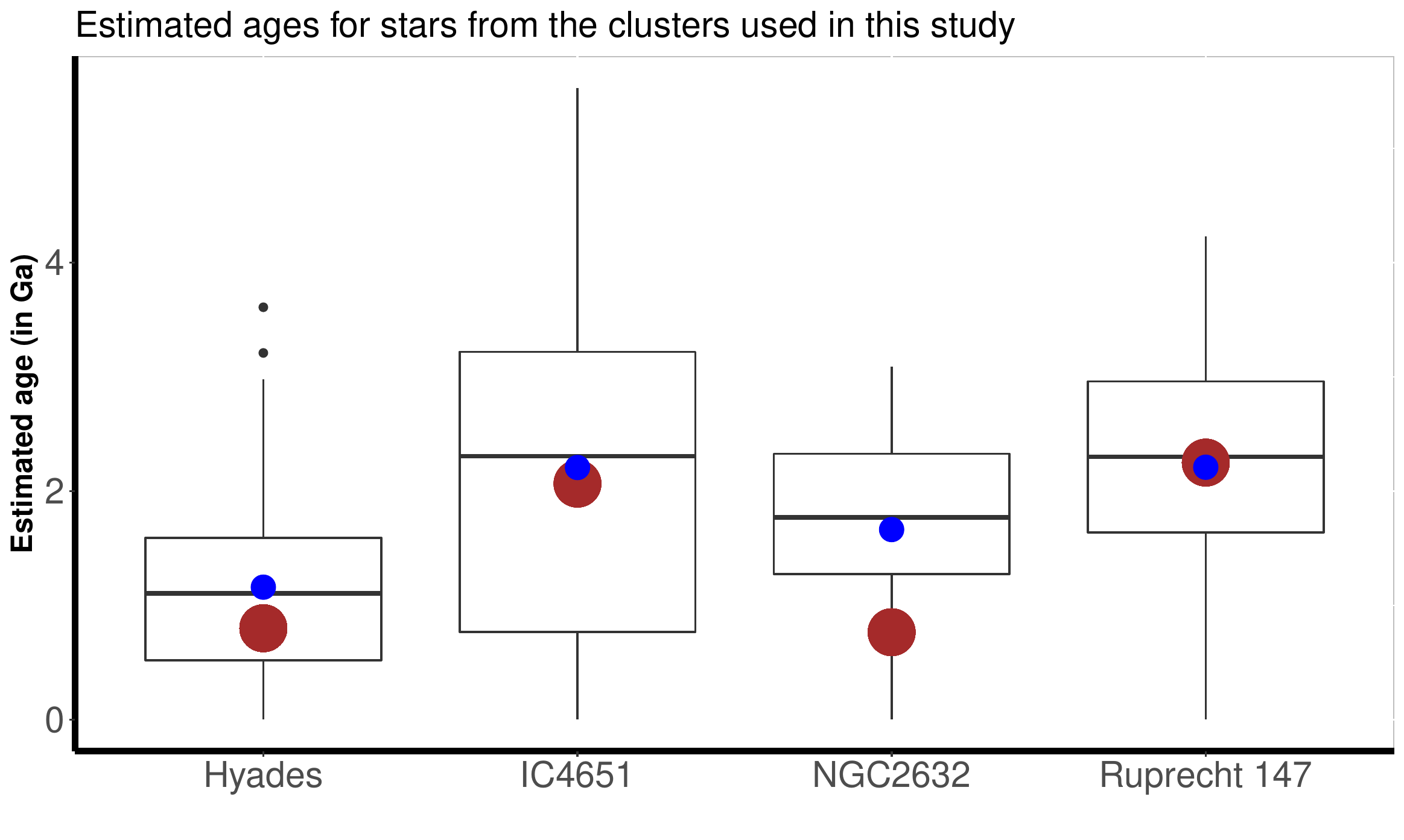}
\caption{Boxplots of the estimated age distribution for the clusters under study using our HBM. The red dots represent the literature ages of the clusters, and the blue dots show the mean ages we obtained.}\label{clusters_pred}
\end{center}
\end{figure}

For this study, we used the data of \citet{Casamiquela}, who analysed the open clusters Hyades, NGC 2632, and Ruprecht 147. They obtained $T_{\rm eff}$, $\log g$, [Fe/H], and the abundances of [Mg/Fe], [Al/Fe], [Ti/Fe], [Si/Fe], and [Y/Fe] for groups of stars in each cluster. In total, 58 stars belong to the Hyades, 18 belong to NGC 2632, and 17 belong to Ruprecht 147. The galactocentric radii of these three clusters are larger than 7 kpc \citep{Wu09,McMillan13}, out of the inner disk, where \citet{Casali} found that these CCs do not work correctly. We also added ten stars from the cluster IC4651 studied by \citet{Blanco}. The galactocentric radius of this cluster is also larger than 7 kpc. For each cluster we identified the outliers, if any. In the case of Hyades, we found four outliers, one for Ruprecht 147, and none for IC4651 and NGC2632. The quantity of outliers is related to the population of stars belonging to each cluster in the testing set. These five outliers were removed for the following analysis and are studied separately in Sec. \ref{outliers}.

In Fig. \ref{clusters_pred} we provide boxplots of the distribution of estimated ages obtained for four clusters. Each boxplot is represented by a rectangle bounding the first and third quartiles of the distribution. The thick black line represents the median. The thin vertical lines bound the maximum and minimum values of the distribution and empty circles represent the outliers. In addition, we included blue points showing the mean of each distribution, and a brown point for the assumed age of the cluster. Our model estimates a mean age for NGC 2632 and Hyades that is slightly older than the assumed age, with a difference between the estimated mean age and the age from the literature of 0.36 Ga and 0.9 Ga for Hyades and NGC 2632, respectively. In the case of Ruprecht 147, the mean value of our estimated ages is 0.04 Ga younger than the assumed age of this cluster. IC4651 presents a mean age 0.14 Ga older than the accepted age for the cluster. The MADs for these clusters are 0.712, 1.35, 1.02, and 0.95 Ga for Hyades, IC4651, NGC2632, and Ruprecht 147, respectively.

This means that although there is a bias larger than expected for the younger clusters, especially for NGC2631, most of the stars in general present age estimates using CCs with accuracies of about 1 Ga, confirming the results for field stars with well-known ages. This is a statement regarding a statistical behaviour, however. There are individual outliers in some of these clusters with estimates even higher than 5 Ga for a few stars.
A summary of the results of all these testing sets (Sections \ref{23_buenas}, \ref{Spina}, \ref{Gaia}, and \ref{Clusters}) can be found in Table \ref{gen_results}.

\begin{table}
        \caption{Summary of the general results obtained with our HBM in comparison to the test ages for all the testing sets. MD accounts for the mean differences, S.D. for the standard deviation of these differences, and MAD for the mean absolute differences. {\it Asteroseismic + others} represents the stars studied in Sec. \ref{23_buenas}, \citet{Spina18} those studied in Sec. \ref{Spina}, {\it Gaia Benchmark} to those in Sec. \ref{Gaia}, {\it Casamiquela} to the clusters taken from \citet{Casamiquela}, and {\it Blanco} to the cluster ID4651 taken from \citet{Blanco}.}
        \label{gen_results}
        \centering
        \begin{tabular}{cccc}
                \hline
                Testing group & MD & S.D. & MAD \\
                Blanco & -0.142 & 1.76 & 1.35 \\
                Casamiquela & -0.397 & 0.98 & 0.818 \\
                Gaia Benchmark & 0.22 & 1.44 & 0.86\\
                Spina et al., 2018 & 0.38 & 1.12 & 0.93\\
                Asteroseismic + others & 0.19 & 1.21 & 0.86\\
                \hline
        \end{tabular}
\end{table}

\subsection{Outlier analysis}\label{outliers}

\begin{figure}
\begin{center}
\includegraphics[width=\linewidth]{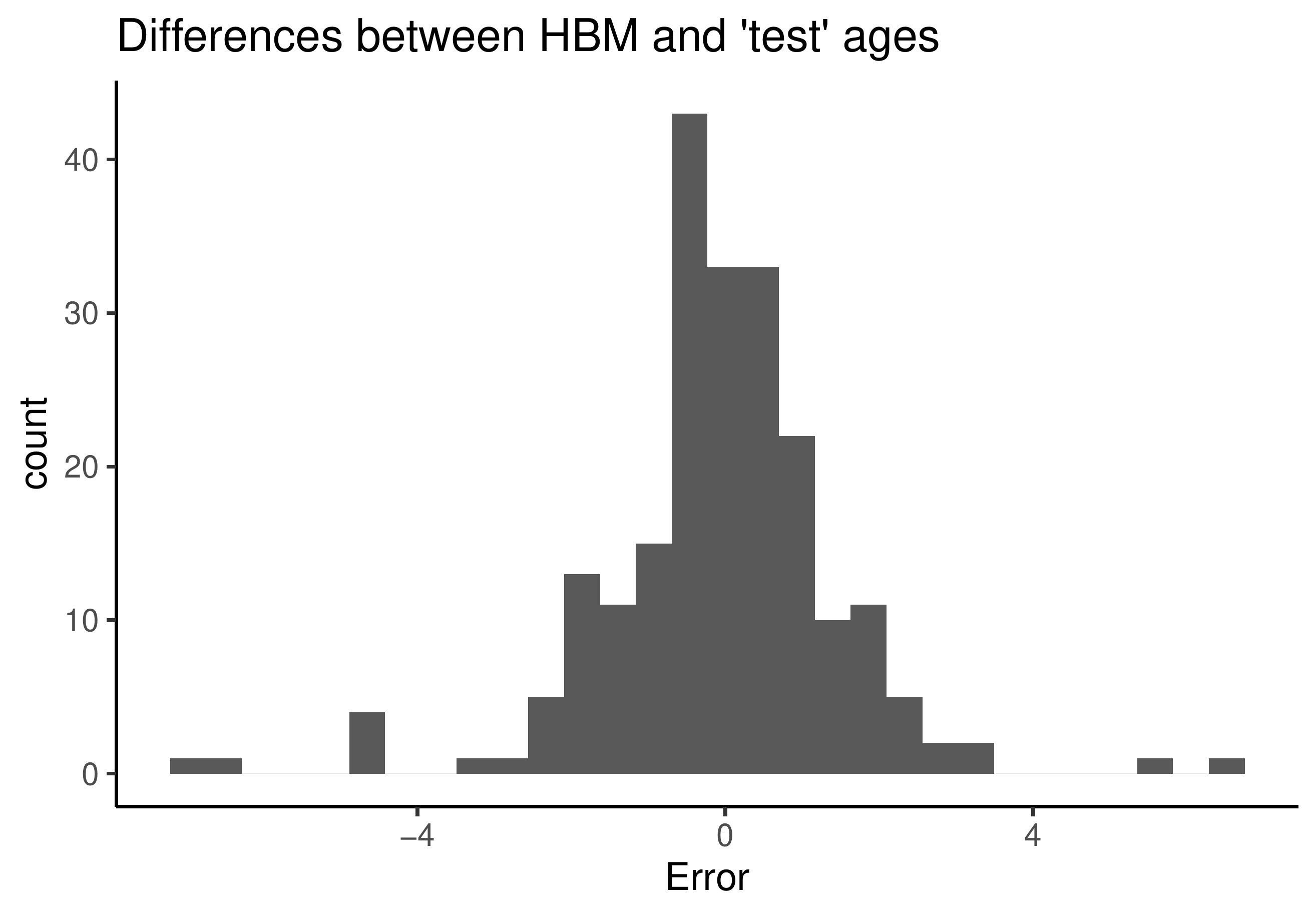}
\caption{Histogram of the differences between the ages estimated by our model and the 'test' ages for the 215 stars of out testing set.}\label{hist_errors}
\end{center}
\end{figure}

In Fig. \ref{hist_errors} we show a histogram with all the differences between the ages estimated using our HBM and the test ages from the literature. This histogram shows eight stars whose inaccuracies clearly exceed those of the main group around Error=0, all higher than 4 Ga in absolute value. These eight stars are those identified in Sec. \ref{Spina} and \ref{Clusters}, and $\alpha$ Cen B and $\tau$ Cet from the Gaia benchmark sample. In addition, we identified four additional stars to be studied in this section: three stars from Sec \ref{23_buenas} that are not outliers when all the testing stars are considered, but they are outliers when only these 23 stars are considered, and the special case of $\epsilon$ Eri. Thus we identified 12 outliers from a total of 215 stars, that is, 5.6 $\%$ of the complete testing sample. All these stars were removed from the previous analysis and are studied in detail in this section.

The outlier found in the set of \citet{Spina18}, as we commented before, was also found and studied by these authors. They concluded that the chemical peculiarities of this star deserve a dedicated analysis that it is beyond the scope of this work. We refer to \citet{Spina18} for the details of this case. The most interesting point here is that we found more cases like this in terms of chemical peculiarities in the other samples we studied.

\subsubsection{$\alpha$ Cen B}

This is one of our most worrying outliers. \citet{Sahlholdt} estimated an age of 5.5$\pm$1.5 Ga, and our HBM estimates an age of 0.11-0.02+2.1 Ga. Our estimate seems unreasonable, but considering the different estimates shown in \citet{Sahlholdt}, this star is quite pathological. It is part of a triple system and should have a similar age as $\alpha$ Cen A, but the literature disagrees about this. The asteroseismic estimate of \citet{Lundkvist} points to a very young star, like  our results; gyrochronology in general points to a star older than $\alpha$ Cen A, and all the estimates range between 1 and 9 Ga. Our estimated uncertainties overlap with the youngest estimates for this star. Nevertheless, there may be a reason for our low estimate. With a $T_{\rm eff}$ in the lower 2.5$\%$ range of our training sample, a $\log g$ in the top 2.5$\%$, and [Sr/Mg] pointing to a very young star, the results are not as reliable as the remaining cases that are properly covered by the training sample.

\subsubsection{$\tau$ Cet}

\begin{figure}
\begin{center}
\includegraphics[width=\linewidth]{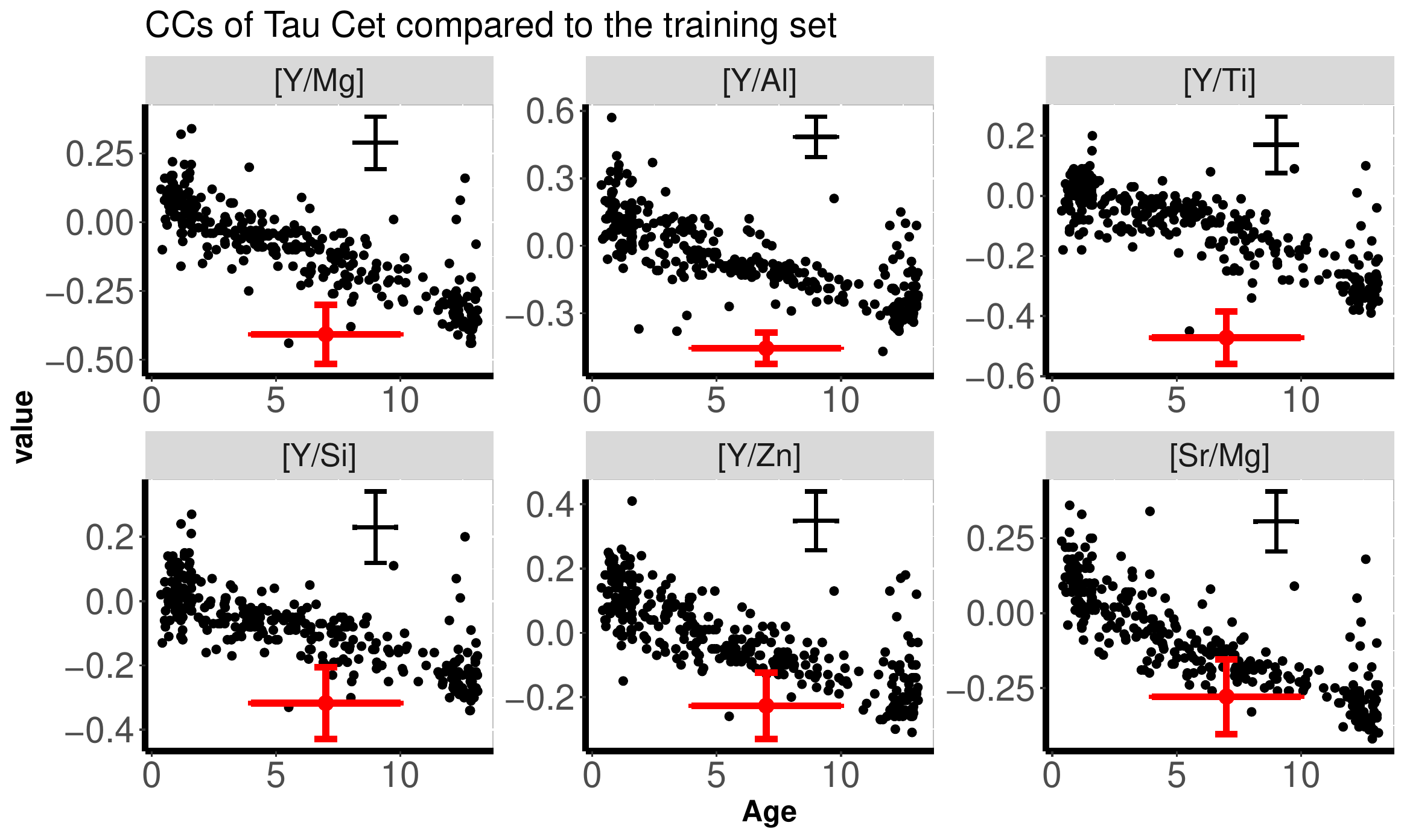}
\caption{Position of the CCs of $\tau$ Cet in the CCs vs. age diagrams. Red dots are the CCs of this star. Black dots are the CCs of the training sample. The black cross represents the mean $1\sigma$ uncertainty of the training sample.}\label{tau_cet}
\end{center}
\end{figure}

This is one of our outliers with clearer chemical peculiarities or an incorrect dating in the literature. The estimates for $\tau$ Cet  in the literature cover a wide range (0 - 14 Ga), with large uncertainties. Therefore, \citet{Sahlholdt} listed an age of 7$\pm$4, that is, in the middle of the range of estimates in the literature. However, their own analysis points to a very old star. Our estimate is 14+0-1.2 Ga, supporting the analysis of \citet{Sahlholdt}. Its CCs  are all compatible with an age older than 10 Ga, with some lying at the top limit of the training sample (the older stars), for example, [Y/Ti], [Y/Al], and [Y/Si]. All this is shown in Fig. \ref{tau_cet}, where the position of the red dots is clearly different from those corresponding to its age in the literature compared to the training sample. They point to an older star. We repeat that this is an indication of chemical peculiarities for this star or an incorrect dating in the literature.

\subsubsection{Other stars with chemical peculiarities}

\begin{figure}
\begin{center}
\includegraphics[width=\linewidth]{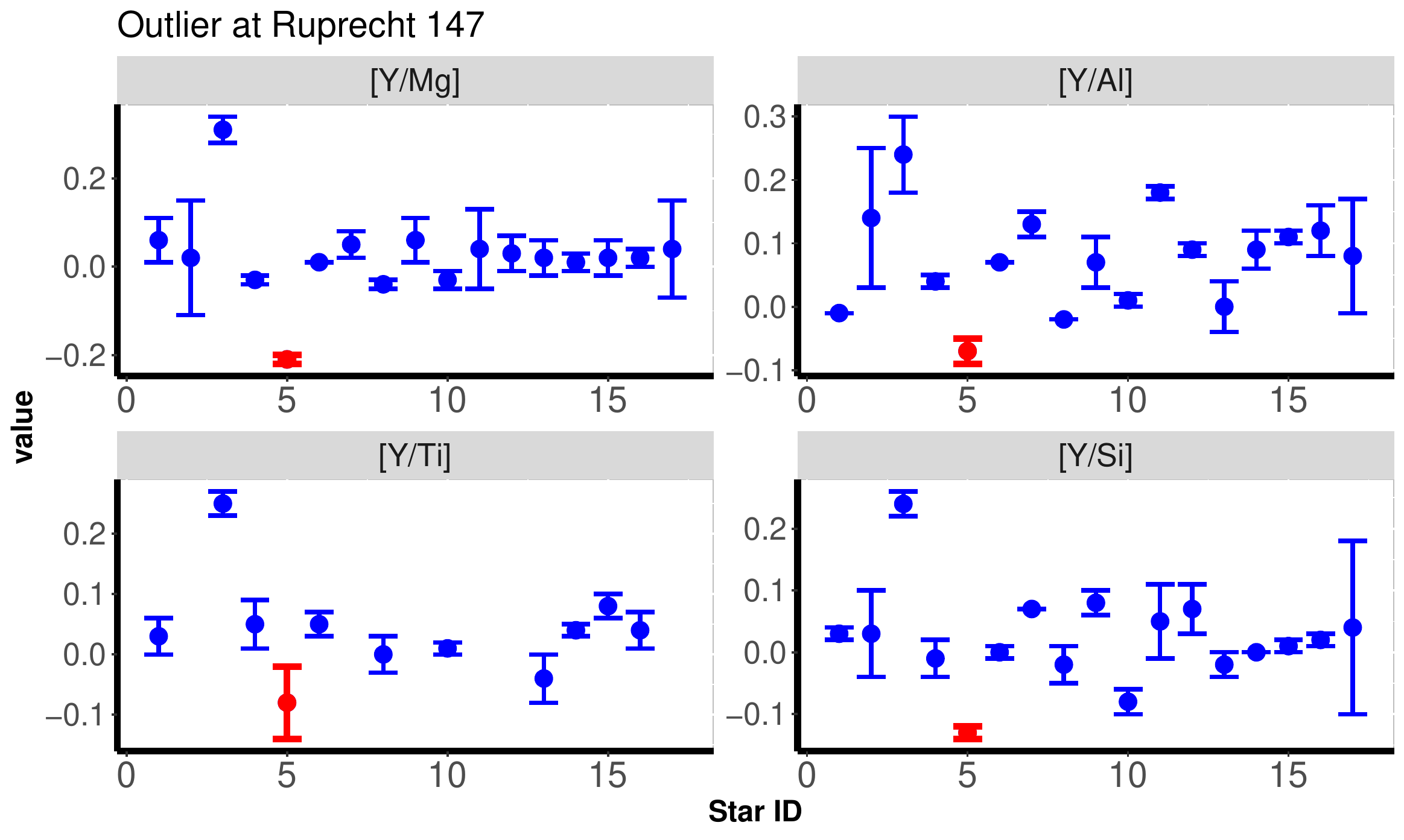}
\caption{CCs of the stars belonging the open cluster Ruprecht 147. The red points represent the outlier found in Fig. \ref{hist_errors} for this open cluster. The x-axis represents the star ID of our sample. The red point corresponds to the star Gaia DR2 4087853875535923200.}\label{r147_out}
\end{center}
\end{figure}

Therefore, we can use CCs in combination with other stellar dating methods to identify potentially very interesting cases with chemical peculiarities. Another example of chemical peculiarities is the only outlier found in Ruprecht 147. In Fig \ref{r147_out} we highlight (with a red symbol) the values of the CCs of this star. Its chemical peculiarities clearly differ from the general trend of the remaining stars in this cluster. The values of the CCs for this outlier point to an older star compared to the rest of its companions. Our model only reflects this fact. Distinguishing the origin of these peculiarities is beyond the scope of this work, but two possible explanations for the outliers found in clusters may be the real chemical peculiarities of these stars, or stars that do not really belong to the cluster. The remaining outliers in clusters, in addition to KIC3656476 and KIC6603624, have a similar behaviour.

\subsubsection{Special cases of KIC8006161 and $\epsilon$ Eri}

In these two cases we found a intrinsic variability in the age estimate of our HBM when we reran it. The reason is related to the combination of two factors: a relatively low number of sampling points (3000, see the appendix), and inconsistent input variables or variables that are not correctly covered by our training sample. These inconsistencies are illustrated using $\epsilon$ Eri as an example. \citet{Sahlholdt} estimated an age of 0.65$\pm$0.25 Ga, and our HBM estimates an age of 4.1-2+2.1 Ga. In this case, gyrochronology and chromochronology provide very stable estimates between 0 and 1 Ga. On the other hand, isochrone fitting provides highly variable results, with ages up to 14 Ga. In a recent work, \citet{Petit21} studied this star using data from SPIRou, NARVAL, and TESS, confirming its young age by the presence of a debris disk.

\begin{figure}
\begin{center}
\includegraphics[width=\linewidth]{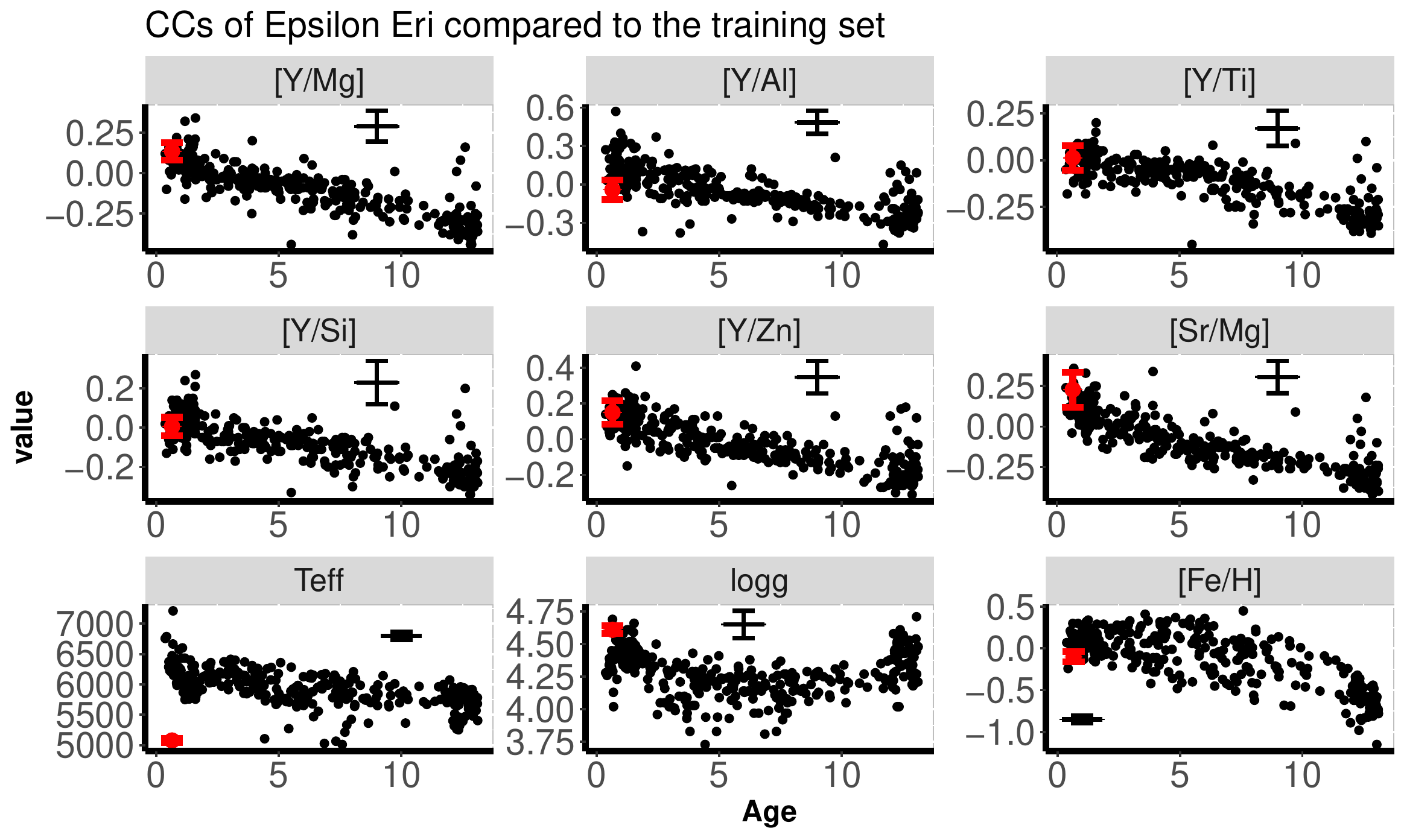}
\caption{Input variables of $\epsilon$ Eri. The red points represent the values for this star. The black points show the corresponding values of our training sample. The black cross represents the mean $1\sigma$ uncertainty of the training sample.}\label{eps_eri}
\end{center}
\end{figure}

In Fig. \ref{eps_eri} we compare the values of the input variables of this star with those of our training sample. In this case, [Fe/H] and [Y/Al] mildly indicate an older star. When these ratios are dismissed, our estimate is reduced in 1 Ga, overlapping the estimate of \citet{Sahlholdt}. We must take into account, however, that the $T_{\rm eff}$ of this star is in the lower 2.5$\%$ of our training sample, making our result not as reliable as for the remaining cases that are properly covered by the sample. This lack of reliability is confirmed with the variability we found when we ran the HBM several times on this star. Therefore, as we described in Sec. \ref{sec:hbm}, we ran this HBM ten times on each star. The final result we used for the comparisons is the mean age from these ten estimations. The reason for this procedure is located in the behaviour of this star and KIC8006161. If we calculate the standard deviation of these ten realisations for $\epsilon$ Eri, we obtain an S.D. of 1.46, and for KIC8006161, the S.D. is 0.76. These are very high values.

As we mentioned, the combination of input values pointing to different ages and, in some cases, with values poorly covered by our training sample, with only 3000 sampling points generates this variability due to the stochastic nature of the estimation algorithm. If we increase the number of sample points to 30000, for example, we find that the mean age estimate remains unaltered (from 3.16 Ga to 2.95 Ga in the case of $\epsilon$ Eri, e.g.), but the S.D. of the ten realisations decreases from 1.47 to 0.65. Nevertheless, we decided to keep this low number of sampling points for the inference because the final age estimate is not affected, but this dispersion offers a quality test for our estimates.

\begin{figure}
\begin{center}
\includegraphics[width=\linewidth]{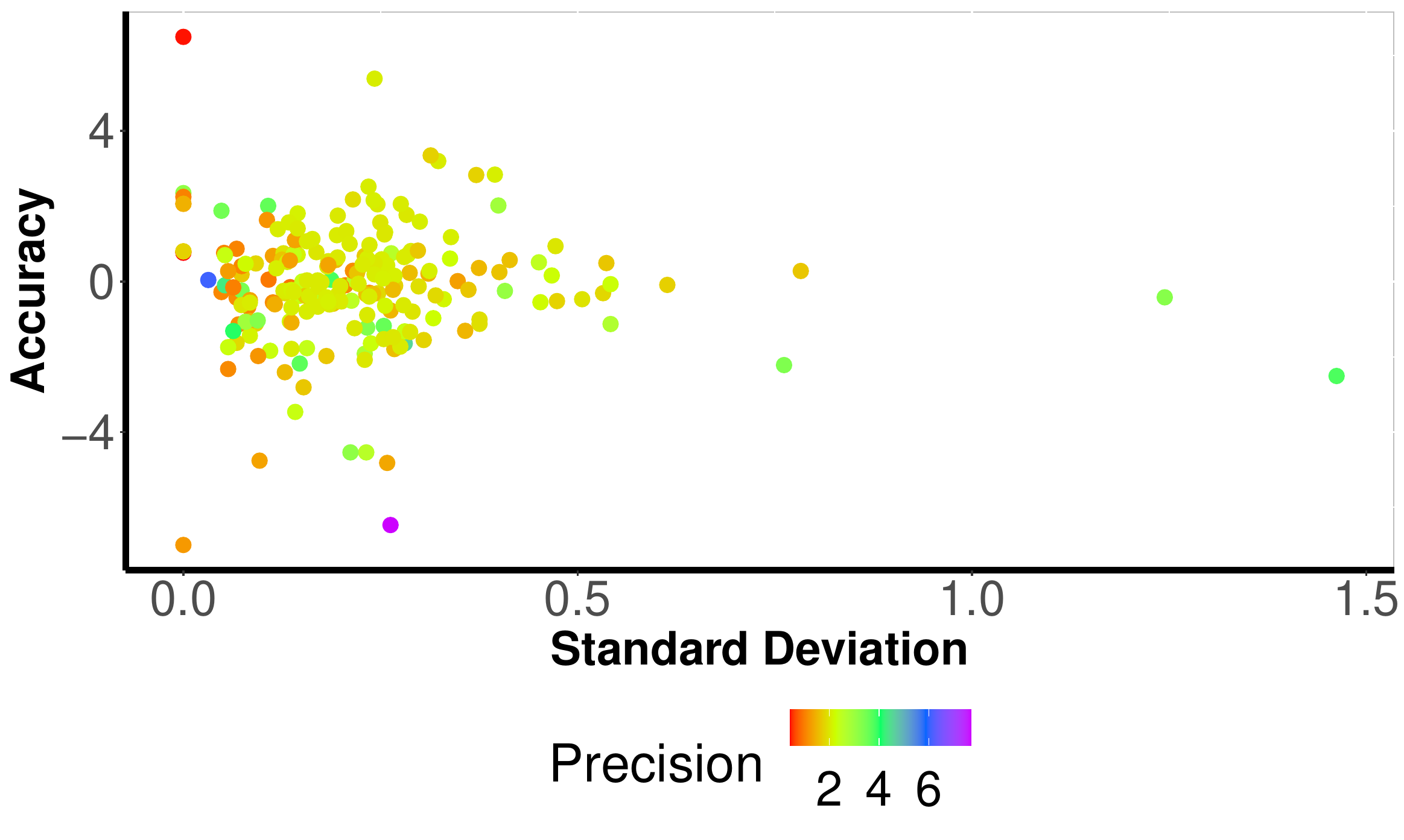}
\caption{Accuracy of our predictions (observed - estimated age) vs. the standard deviation of the ten realisations. The $1\sigma$ uncertainty (precision) is shown in colour.}\label{sd_error}
\end{center}
\end{figure}

This is shown in Fig. \ref{sd_error}, where we show the accuracy of our predictions (or the differences between the HBM estimations and the test ages) versus the standard deviation of these ten realisations. The precision, or $1\sigma$ uncertainty range, is shown in colour. The vast majority of the stars present a low S.D. and reasonable accuracy. On the other hand, we identify two ranges of extreme values:

\begin{itemize}
    \item Those with low S.D. and poor accuracy. These stars belong to the outliers described at the begining of this section. The inputs are consistent, and the age differences are only due to physical reasons.
    \item Those with high S.D. (about five stars). Here the S.D. is a signal of an inconsistency in the input variables, a poor coverage from the training sample, or both. This high S.D. shows that the age estimate provided by the HBM is not reliable and must be taken with caution.
\end{itemize}

In both cases, a poor precision can be indicative of inputs with large uncertainties, and the result must also be taken with caution. In general, a high S.D. is also followed by a precision worst than 3 Ga.

\section{HBM using only CCs}

In general, in the literature CCs are used alone or in combination of the stellar metallicty for stellar dating. In the model we presented, we added information from $T_{\rm eff}$ and $\log g$ for reasons already exposed, but the question is how these results compare with using only CCs and [Fe/H]. To answer it, we trained an HBM only using CCs and [Fe/H]. In this section we present the comparison, using our testing field stars, between the results obtained with this simplified model.

\begin{figure}
\begin{center}
\includegraphics[width=\linewidth]{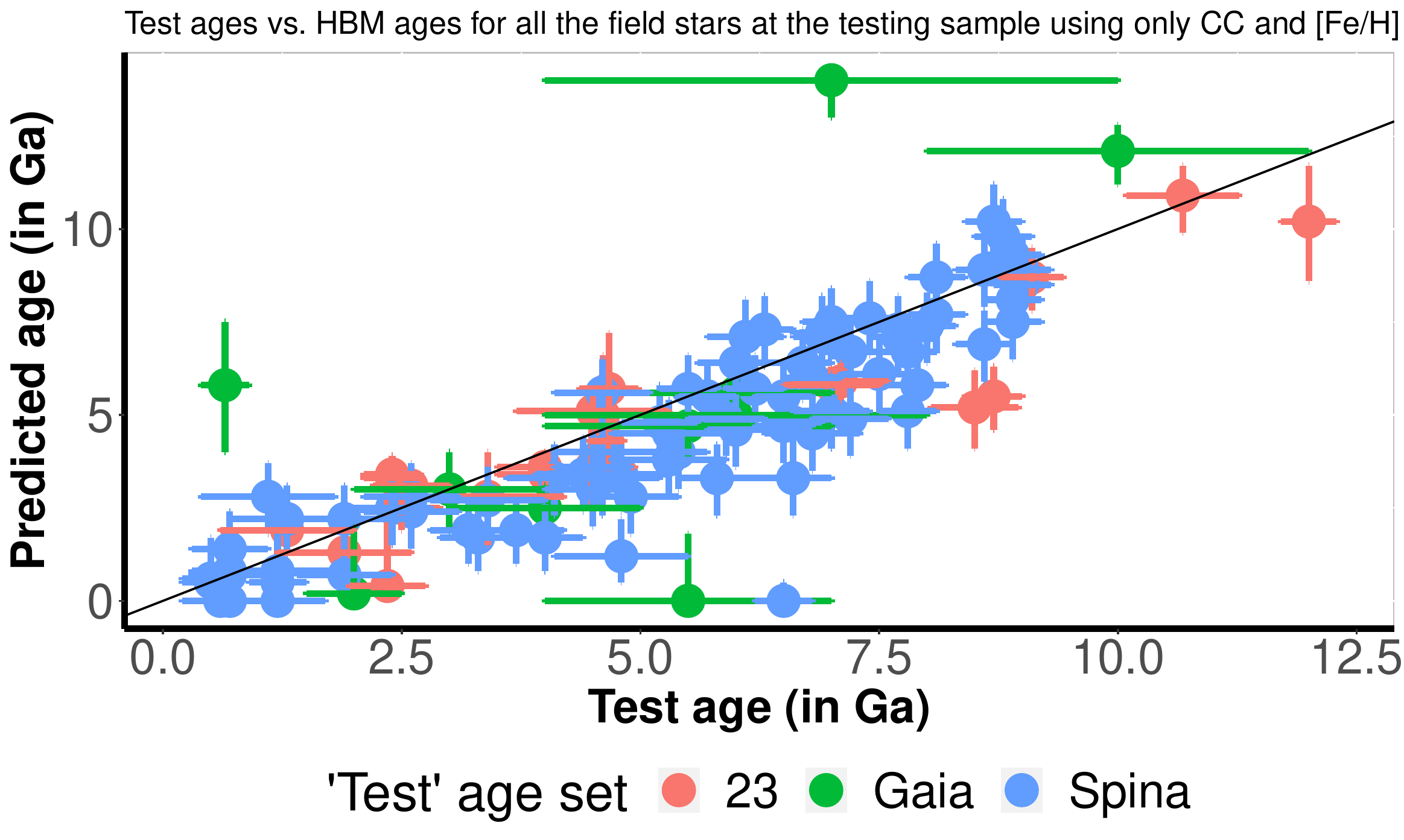}
\caption{Age predictions for the stars studied in sections \ref{23_buenas}, \ref{Spina}, and \ref{Gaia} using our HBM with only CCs and [Fe/H].}\label{todos_juntos_solo_CC}
\end{center}
\end{figure}

In Fig. \ref{todos_juntos_solo_CC} we show the equivalent to Fig. \ref{all_tog}, but this model used only CCs and [Fe/H]. The result, as expected, is very similar to the result obtained using all the input variables, but with a larger dispersion and some new outliers. This is confirmed in Table \ref{solo_CC}. Compared to Table \ref{gen_results}, the bias (MD) in general increases and the MAD does as well, although this increase is not significant in some subsets. This means that this simplified model can be used for a first estimate, avoiding some of the inconsistencies we have found for peculiar stars, but for the best possible accurate age estimate, it is better to use all the input variables.

\begin{table}
        \caption{Statistical results for only CCs and [Fe/H] as input variables, to be compared with Table \ref{gen_results}.}
        \label{solo_CC}
        \centering
        \begin{tabular}{cccc}
                \hline
                Testing group & MD & S.D. & MAD \\
                Blanco & -0.155 & 1.96 & 1.58 \\
                Casamiquela & -1.06 & 1.29 & 1.36 \\
                Gaia Benchmark & 0.486 & 1.29 & 1.09\\
                Spina et al., 2018 & 0.673 & 1.08 & 1.00\\
                Asteroseismic + others & 0.515 & 1.20 & 0.971\\
                \hline
        \end{tabular}
\end{table}

\section{Conclusions}

The use of stellar chemical abundances for stellar dating using certain abundance ratios (the chemical clocks, CCs) has been improved in recent years with the definition of more than ten CCs, and the extension of the use of CCs beyond solar twins.
We took advantage of the exceptional database presented in DM19 to go one step further in the use of CCs for stellar dating. We trained a hierarchical Bayesian model to combine the information coming from different CCs and other physical observables such as $T_{\rm eff}$, $\log g$, or [Fe/H] to provide robust stellar age estimates with good precision.

To test our model, we gathered a number of different testing sets. We used stars with ages estimated by asteroseismology, Gaia benchmark stars, stars from other studies in the context of CCs, and four stellar clusters. We found that our estimates using CCs and an HBM are similar to the reference ages for almost all the tested stars. Compared with all the testing samples, our estimates present an MAD of about 0.91 Ga, with a really short MD of 0.008 Ga, which reflectsthe error compensations. A most reliable MD was obtained by studying the individual testing subsamples, where MD ranges between -0.4 and 0.4 Ga.

Nevertheless, we must take some important aspects related to this technique into account. It is based on the statistical properties of a large amount of stars. Therefore, its predictions for individual stars must always be taken with caution because we are not safe from outliers related to chemical peculiarities or stellar parameters, or observed CCs that are poorly covered by our training sample. Nevertheless, our test shows that the estimates we provide are generally very good age indicators. If this technique is used in combination with any other age estimator, outliers can be identified, and then very interesting cases can be discovered from the point of view of a chemical abundance. In addition, we verified that our model is sometimes not as accurate for cool stars (below 5200 K) as it is for Sun-like stars. For these cool stars, abundance determinations are not completely reliable in our training sample. In general, we suggest that our method is used to predict ages only for stars with stellar characteristics within the 95$\%$ ranges of the properties of the training sample, as shown in Section 2, in order to avoid boundary effects and hence inaccuracies in estimating stellar ages with our model.

We have also presented a simplified version of the HBM using only [Fe/H] and CCs. This model provides reasonable and useful age estimates, but with lower accuracies and precisions than the estimates obtained with the complete model.

One of the main benefits of this dating technique is that it is almost independent of the stellar structure and evolution. It depends mainly on the chemical evolution of certain parts of the Galaxy, where the training sample is located, and the chemical abundances of the original cloud from which the stars formed. Finally, we note that while the ages of the training sample we used were obtained using isochrone fitting, it seems that this has a small impact on the final results.

\begin{acknowledgements}
The authors want to thank an anonymous referee for his/her very interesting and constructive comments. The paper was clearly improved thanks to them. We also want to thank editor M. Salaris for his understanding during the refereeing procedure. AM acknowledges funding support from the European Union's Horizon 2020 research and innovation program under the Marie Sklodowska-Curie grant agreement No 749962 (project THOT), from Grant PID2019-107061GB-C65 funded by MCIN/AEI/ 10.13039/501100011033, and from Generalitat Valenciana in the frame of the GenT Project CIDEGENT/2020/036. V.A. and E.D.M. were supported by FCT - Funda\c{c}\~ao para a Ci\^encia e Tecnologia (FCT) through national funds and by FEDER through COMPETE2020 - Programa Operacional Competitividade e Internacionaliza\c{c}\~ao by these grants: UID/FIS/04434/2019; UIDB/04434/2020; UIDP/04434/2020; PTDC/FIS-AST/32113/2017 \& POCI-01-0145-FEDER-032113; PTDC/FIS-AST/28953/2017 \& POCI-01-0145-FEDER-028953. V.A. and E.D.M also acknowledge the support from FCT through Investigador FCT contracts nr.  IF/00650/2015/CP1273/CT0001 and IF/00849/2015/CP1273/CT0003, and POPH/FSE (EC) by FEDER funding through the program ``Programa Operacional de Factores de Competitividade - COMPETE''.
\end{acknowledgements}


\newpage
\begin{appendix}
\section{HBM analysis}

In this appendix we show all the tests we performed to ensure the consistency of the HBM we used in this study. The main characteristics of the final selected model can be found in Table \ref{mod_carac}. The number of samples, burn-in iterations, and independent chains is considered sufficient to ensure convergence and representativity of the chains. The multivariate Gaussian priors are highly non-informative.

\begin{table*}
        \caption{Main characteristics of the HBM model.}
        \label{mod_carac}
        \centering
        \begin{tabular}{cc}
                \hline
                Total number of iterations & 60000 \\
                Burn-in iterations & 30000 \\
                Number of independent chains & 12\\
                Iterations per chain & 5000\\
                Variables in the correlation matrix & $T_{\rm eff}$, $\log g$, and [Fe/H]\\
                Initial prior for standard deviation & Cauchy(0, 5)\\
                Initial prior for correlation matrix & LKJ($\nu=1$)\\
                Number of sample points used for age estimations & 3000\\
                \hline
        \end{tabular}
\end{table*}

\begin{figure}
\begin{center}
\includegraphics[width=\linewidth]{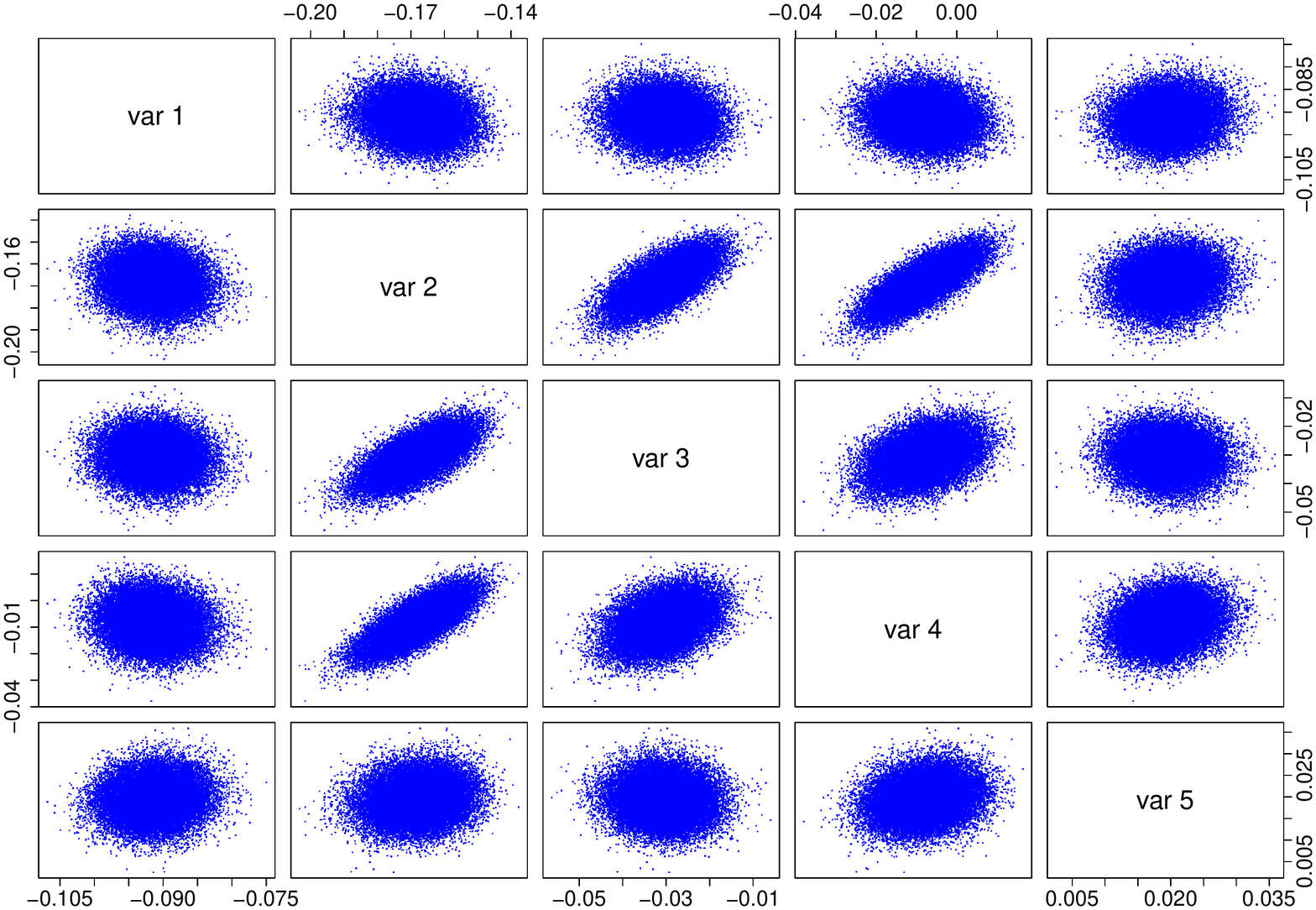}
\caption{Posterior distribution of the coefficients of Eq. \ref{deterministic} in the case of the CC [Y/Mg]. Var 1 to Var 5 correspond to $k_1$ to $k_5$ in this equation.}\label{posteriors}
\end{center}
\end{figure}

The posterior distribution for the different coefficients can be found in Fig. \ref{posteriors} for the case of the relation of one of the CCs, in this case, [Y/Mg]. In this figure, Var 1 to Var 5 represent the coefficients $k_1$ to $k_5$ of Eq. \ref{deterministic}. This figure shows that there are no clear correlations between these posteriors, except for the coefficient accompanying the age with that accompanying [Fe/H] and slightly for that accompanying $T_{\rm eff}$. In any case, the inclusion of these quantities in the correlation matrix ensures the consistency of the model. Another conclusion is that the presence of $\log g$ in this matrix is not imperative because no clear correlations are found. 

The convergence of the MCMC chains can be verified using the $\hat{R}$ parameter \citep{gelmanbda04}. The closer $\hat{R}$ is to one, the better the convergence of the model. This parameter is measured for every variable of the model, that is, the 30 $k_{i,j}$ coefficients, the correlation matrix, the variables standard deviations, and the true values of the variables, adding more than 1000 variables. The mean value of $\hat{R}$ for all these variables is 0.9999, with a standard deviation of 0.0003. In conclusion, the general convergence of the chains is ensured.

We also analysed the number of effective sampling points ($n_{\rm eff}$) for each variable. $k_{i,j}$ coefficients are those with the lower $n_{\rm eff}$ values, but in any case $n_{\rm eff}>4000$. Because our model has 30000 sample points, we used only one point of ten to estimate the stellar age, ensuring the statistical consistency of the estimation and speeding up its computation.

In Table \ref{all_sp} we list the impact on the final general results of using all the 30000 sampling points. This table must be compared with Table \ref{gen_results}, where we used one point of ten, but ten realisations. The differences in the final results are negligible, of the second or third significant figure, confirming our choice for the selected model.

\begin{table}
        \caption{Statistical results in the case of using all sampling points for the age estimates, to be compared with Table \ref{gen_results}.}
        \label{all_sp}
        \centering
        \begin{tabular}{cccc}
                \hline
                Testing group & MD & S.D. & MAD \\
                Blanco & -0.185 & 1.75 & 1.32 \\
                Casamiquela & -0.429 & 1.00 & 0.849 \\
                Gaia Benchmark & 0.106 & 1.43 & 1.08\\
                Spina et al., 2018 & 0.332 & 1.12 & 0.921\\
                Asteroseismic + others & 0.450 & 1.23 & 0.923\\
                \hline
        \end{tabular}
\end{table}

A final consistency check is related to the inclusion or exclusion of $\log g$ in the correlation matrix. In Table \ref{correls} we show the Spearman coefficient to understand the correlations between the different independent variables of Eq. \ref{deterministic}. $\log g$ is the variable with the weakest correlations with the other variables. This can also been confirmed with the posterior distribution of Var 5 (the $k_5$ coefficient campaigning $\log g$) in Fig \ref{posteriors}.

\begin{table}[ht]
        \caption{Spearman coefficient for the correlation between the different independent variables of Eq. \ref{deterministic}.}
        \label{correls}
        \centering
        \begin{tabular}{rrrrr}
                \hline
                & Age & $T_{\rm eff}$ & [Fe/H] & $\log g$ \\ 
                \hline
                Age & 1.00 & -0.66 & -0.68 & -0.20 \\ 
                $T_{\rm eff}$ & -0.66 & 1.00 & 0.15 & 0.11 \\ 
                ${\rm [Fe/H]}$ & -0.68 & 0.15 & 1.00 & -0.11 \\ 
                $\log g$ & -0.20 & 0.11 & -0.11 & 1.00 \\ 
                \hline
        \end{tabular}
\end{table}

\begin{figure}
\begin{center}
\includegraphics[width=\linewidth]{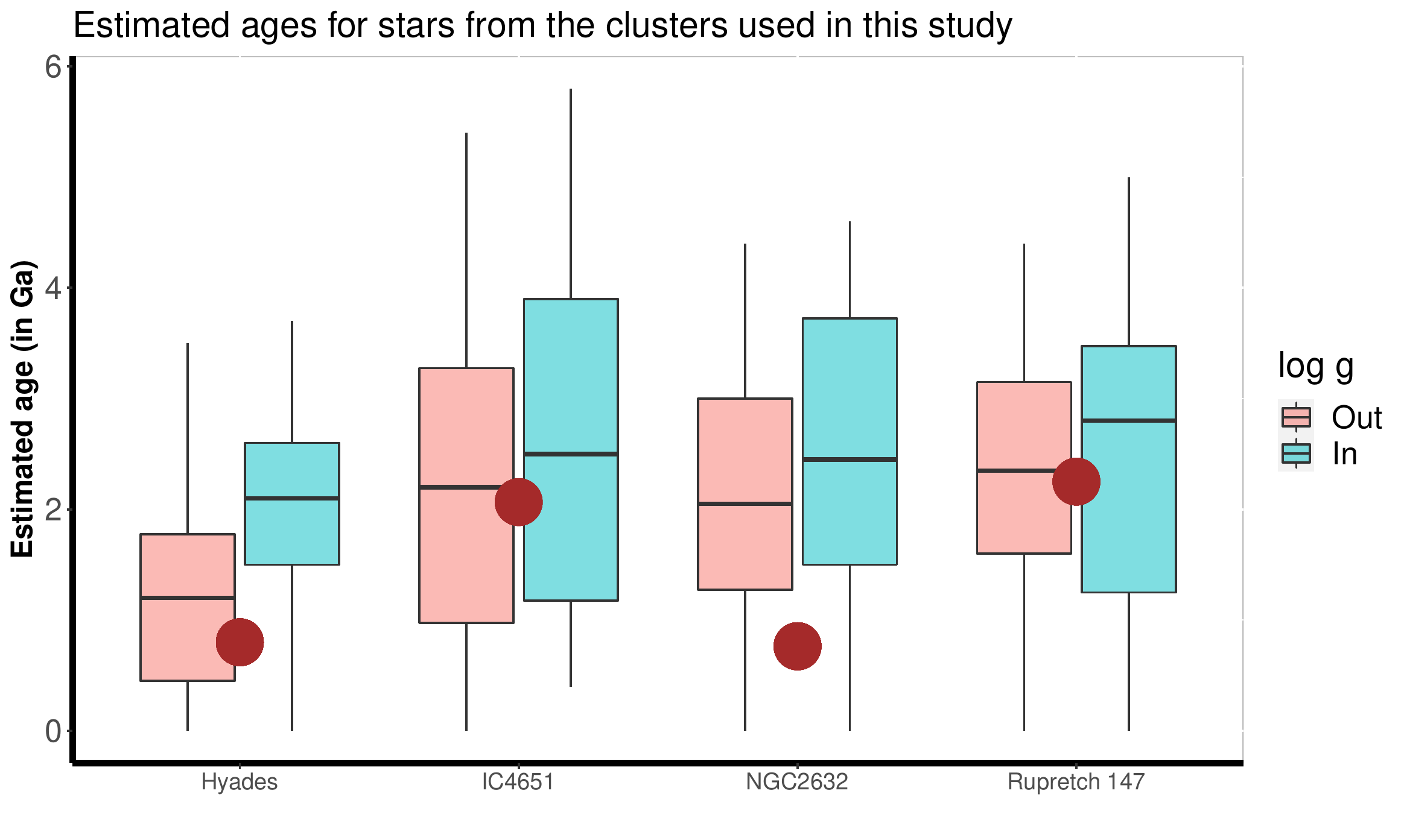}
\caption{Boxplot of the age estimates obtained for all the stars in clusters performed with $\log g$ in the correlation matrix (blue) and out of the correlation matrix (red). Brown points represent the accepted age of each cluster.}\label{clusters_logg_dentro_fuera}
\end{center}
\end{figure}

Therefore, we expect a weak or even negligible impact on the final model and its estimates of including $\log g$ in the correlation matrix. In any case, we performed the test, and in Table \ref{logg_dentro}, we list the final general results when $\log g$ is included in the correlation matrix, to be compared with Table \ref{gen_results}, where $\log g$ is not included. This comparison shows some interesting results. The tests made using field stars show similar results regardless of whether $\log g$ is included. In particular, MD with $\log g$ in the correlation matrix is slightly lower. When we compared stars in clusters (Blanco and Casamiquela sets), we were surprised, however. In these cases, including $\log g$ in the matrix significantly worsens the results. This is illustrated in Fig. \ref{clusters_logg_dentro_fuera}, where we show the boxplot of the results obtained with both options for $\log g$ (inside the correlation matrix in blue and outside this matrix in red). Brown points represent the accepted age of each cluster. In all cases, including $\log g$ provides the worst estimates.

\begin{table}
        \caption{Statistical results in the case of including $\log g$ in the correlation matrix for the age estimates, to be compared with Table \ref{gen_results}.}
        \label{logg_dentro}
        \centering
        \begin{tabular}{cccc}
                \hline
                Testing group & MD & S.D. & MAD \\
                Blanco & -0.575 & 1.81 & 1.47 \\
                Casamiquela & -1.06 & 1.18 & 1.38 \\
                Gaia Benchmark & 0.243 & 1.304& 0.957\\
                Spina et al., 2018 & -0.274 & 1.14 & 0.921\\
                Asteroseismic + others & 0.133 & 1.50 & 1.18\\
                \hline
        \end{tabular}
\end{table}

\begin{table}[ht]
        \caption{Pearson coefficient for the correlation between the different independent variables of Eq. \ref{deterministic}.}
        \label{correls_pearson}
        \centering
        \begin{tabular}{rrrrr}
                \hline
                & Age & $T_{\rm eff}$ & [Fe/H] & $\log g$ \\ 
                \hline
                Age & 1.00 & -0.60 & -0.76 & -0.09 \\ 
                $T_{\rm eff}$ & -0.60 & 1.00 & 0.19 & 0.15 \\ 
                ${\rm [Fe/H]}$ & -0.76 & 0.19 & 1.00 & -0.14 \\ 
                $\log g$ & -0.09 & 0.11 & -0.14 & 1.00 \\ 
                \hline
        \end{tabular}
\end{table}

We can explain this behaviour by recalling that to construct these models, we assumed that correlations between independent variables were Gaussian. Priors were assumed to be multivariable Gaussians. Therefore, if for any reason they were not multivariable, these priors would not be properly supported by the training data. This will have a major impact when these variables are included in the correlation matrix. To identify whether this is the case, we studied the correlations of the independent variables of Eq. \ref{deterministic}, as we did for Table \ref{correls}, but using the Pearson coefficient. This coefficient is based on the assumption of Gaussian correlations, whereas the Spearman coefficient is not. Any significant difference between these two coefficients is an indicator of a non-Gaussian correlation. Comparing Tables \ref{correls} and \ref{correls_pearson}, we find that the differences are lower than 30 $\%$ in general. Only the case of the correlation between age and $\log g$ presents a difference of 128 $\%$. This points to this correlation as the responsible of the differences found when including or excluding $\log g$ in the correlation matrix. This inconsistency is not highly critical for field stars, where ages and $\log g$ are mixed and the low correlation between them has a low impact. In the case of stars in clusters, however, when age is the same for all the members, this inconsistency is enhanced because the Gaussian model inferred from the model is far from the real distribution. As a conclusion, we can say that not including $\log g$ in this matrix does not solve the problem of the inconsistency of the non-Gaussian correlations, but simplifies the model assuming $\log g$ is not correlated to the remaining independent variables (which is a reasonable approximation). This is better than inferring an incorrect correlation. We therefore decided to use the model with $\log g$ out of the correlation matrix.

\end{appendix}

\end{document}